\documentclass[apj]{emulateapj}
\usepackage{apjfonts}

\slugcomment{v4.09 \today}

\newcommand{\msun}{M$_{\odot}$}

\newcommand{\Reff}{{$R_{\rm e}$}}
\newcommand{\asec}{{^{\prime\prime}}}
\newcommand{\amin}{{^{\prime}}}

\newcommand\arraybslash{\let\\\@arraycr}

\begin{document}

\title{The MaNGA Integral Field Unit Fiber Feed System for the
  Sloan 2.5~m Telescope}

\author{N.~Drory\altaffilmark{1},
  N.~MacDonald\altaffilmark{2},
  M.~A.~Bershady\altaffilmark{3},
  K.~Bundy\altaffilmark{4},
  J.~Gunn\altaffilmark{5}
  D.~R.~Law\altaffilmark{6},
  M.~Smith\altaffilmark{3},
  R.~Stoll\altaffilmark{7},
  C.~A.~Tremonti\altaffilmark{3},
  D.~A.~Wake\altaffilmark{3,11},
  R.~Yan\altaffilmark{8},
  A.~M.~Weijmans\altaffilmark{12},
  N.~Byler\altaffilmark{2},
  B.~Cherinka\altaffilmark{6},
  F.~Cope\altaffilmark{9},
  A.~Eigenbrot\altaffilmark{3},
  P.~Harding\altaffilmark{10},
  D.~Holder\altaffilmark{9},
  J.~Huehnerhoff\altaffilmark{9},
  K.~Jaehnig\altaffilmark{3},
  T.~C.~Jansen\altaffilmark{2},
  M.~Klaene\altaffilmark{9},
  A.~M.~Paat\altaffilmark{2},
  J.~Percival\altaffilmark{3},
  and
  C.~Sayres\altaffilmark{2},
}


\altaffiltext{1}{McDonald Observatory, The University of Texas at
  Austin, 1 University Station, Austin, TX 78712, USA}
\email{drory@astro.as.utexas.edu}

\altaffiltext{2}{Department of Astronomy, University of Washington,
  Box 351580 Seattle, WA 98195}

\altaffiltext{3}{Department of Astronomy, University of Wisconsin, 475
  N.\ Charter St., Madison, WI 53706, USA}

\altaffiltext{4}{Kavli Institute for the Physics and Mathematics of
  the Universe (Kavli IPMU, WPI), Todai Institutes for Advanced Study,
  the University of Tokyo, Kashiwa, Japan 277-8583}

\altaffiltext{5}{Department of Astrophysical Sciences, Princeton
  University, Princeton, NJ 08544, USA}

\altaffiltext{6}{Dunlap Institute for Astronomy \& Astrophysics,
  University of Toronto, 50 St. George St, Toronto, ON M5S 3H4, Canada}

\altaffiltext{7}{C~Technologies, Inc., 757 Route 202/206,
  Bridgewater, NJ 08807, USA}

\altaffiltext{8}{Department of Physics and Astronomy, University of
  Kentucky, Lexington, Kentucky, 40506-0055}

\altaffiltext{9}{Apache Point Observatory, P.O.\ Box 59, Sunspot, NM
  88349, USA}

\altaffiltext{10}{Department of Astronomy, Case Western Reserve
  University, Cleveland, Ohio 44106, USA}

\altaffiltext{11}{Department of Physical Sciences, The Open University,
  Milton Keynes, MK7 6AA, UK}

\altaffiltext{12}{School of Physics and Astronomy, University of St
  Andrews, North Haugh, St Andrews, Fife KY16 9SS, UK}

\begin{abstract}
  We describe the design, manufacture, and performance of bare-fiber
  integral field units (IFUs) for the SDSS-IV survey MaNGA (Mapping
  Nearby Galaxies at APO) on the the Sloan 2.5~m telescope at Apache
  Point Observatory (APO). MaNGA is a luminosity-selected
  integral-field spectroscopic survey of 10$^4$ local galaxies
  covering 360-1030~nm at $R\sim2200$.  The IFUs have hexagonal dense
  packing of fibers with packing regularity of 3~$\mu$m (RMS), and
  throughput of 96$\pm$0.5\% from 350~nm to 1~$\mu$m in the lab. Their
  sizes range from 19 to 127 fibers (3-7 hexagonal layers) using
  Polymicro FBP 120:132:150~$\mu$m core:clad:buffer fibers to reach a
  fill fraction of 56\%.  High throughput (and low focal-ratio
  degradation) is achieved by maintaining the fiber cladding and
  buffer intact, ensuring excellent surface polish, and applying a
  multi-layer AR coating of the input and output surfaces. In
  operations on-sky, the IFUs show only an additional 2.3\%
  FRD-related variability in throughput despite repeated mechanical
  stressing during plate plugging (however other losses are present).
  The IFUs achieve on-sky throughput 5\% above the single-fiber feeds
  used in SDSS-III/BOSS, attributable to equivalent performance
  compared to single fibers and additional gains from the AR coating.
  The manufacturing process is geared toward mass-production of
  high-multiplex systems. The low-stress process involves a precision
  ferrule with hexagonal inner shape designed to lead inserted fibers
  to settle in a dense hexagonal pattern. The ferrule inner diameter
  is tapered at progressively shallower angles toward its tip and the
  final 2~mm are straight and only a few micron larger than necessary
  to hold the desired number of fibers. Our IFU manufacturing process
  scales easily to accommodate other fiber sizes and can produce IFUs
  with substantially larger fiber counts. To assure quality, automated
  testing in a simple and inexpensive system enables complete
  characterization of throughput and fiber metrology.  Future
  applications include larger IFUs, higher fill-factors with stripped
  buffer, de-cladding, and lenslet coupling.
\end{abstract}


\section{Introduction}
\label{sec:introduction}

The goal of the integral-field unit (IFU) development project
described here is to enable an environmentally unbiased integral-field
spectroscopic survey of $\sim$10,000 local galaxies selected to
have a roughly flat stellar mass-distribution between
10\textsuperscript{9} and 10\textsuperscript{12}~\msun\ for the
SDSS-IV project MaNGA (Mapping Nearby Galaxies at APO;
\citealp{Bundy+2014}). The survey has no other selection criteria but
luminosity and redshift, and in particular no selection on apparent size
or inclination.  The galaxies are to be sampled with
near-integral spatial coverage to between 1.5 and 2.5 half-light
radii, \Reff, with physical sampling of 1--2~kpc. The spectra are
required to reach a depth equivalent to S/N of $\sim$5 per pixel
($\sim1$~\AA) measured through a 2$\asec$ diameter fiber aperture at
1.5~\Reff.  This enables measurements of gradients in stellar and gas
composition as well as kinematics. A spectral resolution equivalent to
$\sigma\sim60$ km s$^{-1}$ ensures dynamical mass measurements for the
bulk of the sample.  The MaNGA survey and its rationale are described
in detail in \citet{Bundy+2014}.

\begin{figure*}[tb]
  \centering
  \includegraphics[width=0.9\textwidth]{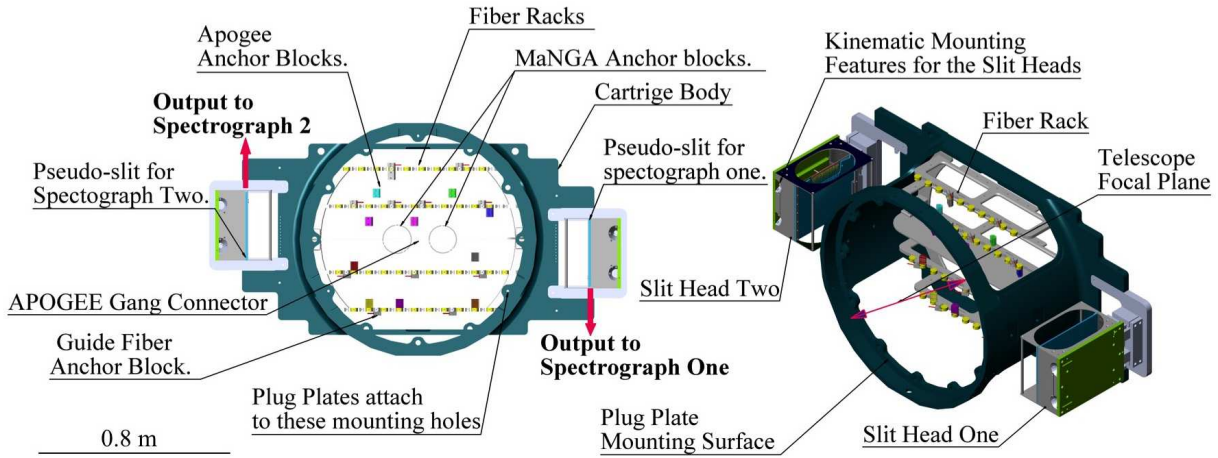}
  \caption{Overview of the Sloan 2.5m Telescope fiber cartridge with
    critical components labeled.  Apogee fibers extend towards the
    plug plate from the APOGEE anchor blocks shown in green.  MaNGA
    fibers extend towards the plug plate from the MaNGA anchor blocks
    shown in read.  Slit heads and cartridge body index to the BOSS
    spectrographs and telescope independently through kinematic
    features on each.\label{fig:cart-overview}}
\end{figure*}

The broader context of the MaNGA IFU development concerns the
cost-performance trades of a purely fiber-based IFUs contrasted to
other approaches to integral-field spectroscopy (IFS) involving
traditional, grating-dispersed spectrographs. Other, more costly
approaches include mirror-based slicers, lenslets, or lenslet-coupled
fibers. Compared to even a decade ago, when optical fibers with
exquisite broad-spectrum transmission in sizes relevant to
astronomical application were widely in use (e.g.,
\citealp{Fabricant+2005}), costs are now lower for high-precision
machining of small components (e.g., wire electric discharge
machining, EDM), and improvements continue in anti-reflection (AR)
coating. These developments enable affordable ways to assemble fibers
into arrays that are nearly perfect in both their metrology and
throughput. When properly handled and fed with suitably fast beams
(within a factor of two of their numerical aperture), entropy
increase--focal ratio degradation (FRD)--is minimal.  While
fiber-based IFUs will never achieve 100\% filling factor without
lenslet coupling or significant sacrifice in throughput and
cross-talk, from a survey perspective truly integral spatial coverage
in a single exposure is not necessarily the salient metric. Further,
with improvements in fiber metrology, mis-alignment losses in lenslet
coupling can be minimized in future instruments where truly integral
coverage is at a premium.

This paper contains a complete description of the design, fabrication,
and the lab as well as on-sky performance of the fiber IFUs used for
MaNGA in SDSS-IV. In the following section (\ref{sec:system-overview})
we provide an overview of the existing spectrograph and fiber
plug-plate systems. This serves as a boundary condition for our IFU
design. Based on this analysis and survey design requirements, in
section \ref{sec:ifu-design} we give a description of the fiber
harnesses, including the custom fiber draw (\ref{fiber-selection}),
ferrule design and fiber mapping (\ref{ferrule-design}), components
and assembly of the fiber pseudo-slit (\ref{sec:v-groove-block} and
\ref{sec:slit-plate-assembly}), AR coating (\ref{sec:ar-coating}) and
cabling strain relief (\ref{sec:strain-relief}).  Section
\ref{metrology} contains a description of the hardware metrology,
including the fiber test-stands developed and used for our analysis.
Section \ref{sec:on-sky-performance} provides a description of
as-built performance. The summary and conclusions are given in section
\ref{sec:conclusions}.

\section{System Overview}
\label{sec:system-overview}

The IFU fiber feed system is designed to couple to the legacy SDSS
hardware including the Sloan 2.5~m telescope and spectrograph corrector
optics \citep{Gunn+2006}, the BOSS spectrographs and
spectrograph-telescope mounting \citep{Smee+2013}, fiber cartridges,
guider system, and plug-plate system \citep{York+2000} at Apache Point
Observatory (APO). The two BOSS spectrographs, each with two spectral
channels, achieve a spectral resolution of $R\equiv
\lambda/\delta\lambda \sim 1900$ in the blue and $\sim 2500$ in the
red using a 2$\asec$ fiber diameter. The resolution varies linearly
with wavelength between 1370 at 360~nm and 2280 at 600~nm in the blue
channel and between 1780 at 600~nm and 2990 at 1050~nm in the red
channel\footnote{There is a degradation of spectral resolution for
  $\lambda>950$~nm due to an increase in optical aberrations, but
  $R>2200$ to the red limit of the spectrograph.}.

\subsection{Existing Cartridge System}
\label{sec:existing-cartridge-system}

SDSS fiber cartridges (Fig.~\ref{fig:cart-overview}) allow positioning
of fibers on astronomical sources by plugging steel ferrules (holding
the fibers) into holes in aluminum plug plates, custom drilled for
every targeted field. Each plate is mounted into a cartridge, plugged
with fibers, and bent to follow the shape of the curved focal plane
over the 3~degree field of view. This plug plate and cartridge
system--existing infrastructure, proven to be cost-effective and
reliable through SDSS-I, II, \& III \citep{York+2000}--is continuing
to be used in SDSS-IV. The existing cartridge systems is remarkably
flexible, serving to map the telescope focal plane to all survey
spectrographs at APO in SDSS-IV, namely BOSS \citep{Smee+2013} and
APOGEE (Wilson et al.\ in prep).

From the focal plane, the fibers extend 0.7~m to anchor blocks
providing support and strain relief to the sheathing protecting the
fibers.  From there, the fibers feeding the BOSS spectrographs
(including MaNGA) extend to two slit heads where they are terminated
on V-groove blocks mounted onto the slit plates. Each slit head, in
turn, inserts into one of the two BOSS spectrographs once the
cartridge is mounted to the telescope by clamping to the Cassegrain
port. The clamps pull the cartridge to 3 kinematic registration points
for accurate placement of the plate to the focal plane.  Similarly,
each slit head is independently registered to the two spectrographs by
a similar mechanism that pulls the slit head into kinematic pads
inside the instruments. Physical limitations of the building that
houses the plugging and racking facility allows for 17 cartridges, any
10 of which can be used on a given night. Six of these cartridges are
populated with MaNGA IFUs, to be prepared with plates and plugged
during daytime for use on the telescope during the following
night. Exchanging a cartridge for another on the telescope can be
completed in $\sim$10~min.  In SDSS-IV, MaNGA shares these six
cartridges with the APOGEE-2 survey at APO. This enable both surveys
to co-observe.

\subsection{Design Considerations}
\label{sec:ifu-design-considerations}

Our IFU design is restricted to bare-fiber systems similar to those
currently found on the WIYN
\citep{Barden+1998,Bershady+2004,Bershady+2005}, Calar Alto 3.5~m
\citep{Kelz+2006}, McDonald 2.7~m \citep{Hill+2008,Fabricius+2008},
and AAO \citep{Croom+2012} telescopes. Lenslet arrays, for example,
while providing truly integral spatial coverage, are difficult to
implement with the existing telescope and spectrograph beam speeds.

The hardware described in this paper overcomes the challenges of
trying to interface an integral field unit (IFU) system to the
existing SDSS survey infrastructure while accommodating the science
requirements laid out in \citet{Yan+2014}. We work within the
constraints of an f/5 telecentric telescope feed over a curved focal
surface and an f/4 spectrograph collimator (values that work well with
the SDSS-I/II/III single-fiber plug system). We maintain the spectral
resolution at $R\sim2200$ (60~km/s), the velocity scale,
$\sqrt{v^2+\sigma^2}$, associated with the lowest-mass galaxies we
want to survey. This corresponds to a fiber size of 120~$\mu$m core
diameter, or 2$\asec$ on sky. This scale is a good compromise between
spatial resolution (given the typical seeing), target density on the
sky, apparent size of the targets, and exposure time per plate for
galaxies at $z < 0.15$. The apparent size distribution of the targets
then motivates a variety of IFU sizes to guarantee most efficient use
of the fibers over the course of the survey. Packing densities with
active-core fill-factors of $>50\%$ within the IFUs are needed to
obtain complete and uniform spatial coverage with a reasonably small
number of sub fiber-diameter telescope dithers \cite{Law+2014}. This
coverage ensures uniform Nyquist-limited PSF reconstruction and
resolution. To achieve such fill-factors, we require a custom draw of
fiber minimizing clad and buffer thickness while maintain high
throughput to 1050~nm. Finally, the available detector real-estate and
crosstalk between neighboring fibers limits the total number of fibers
we can arrange along the slit to $<750$ per spectrograph.

\begin{figure*}[tbh]
  \centering
  \includegraphics[width=0.9\textwidth]{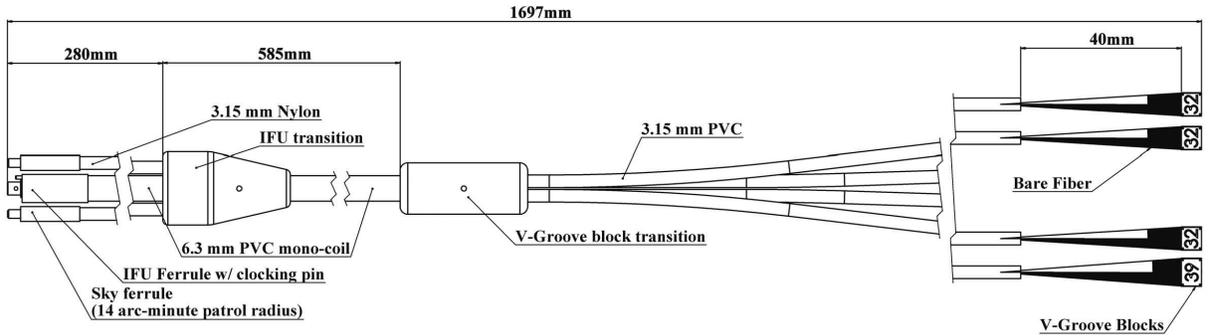}
  \caption{Schematic layout of 127-fiber IFU harness with key
    components labeled.  Only 2 of 8 sky ferrules are shown for
    clarity. The sky ferrules associated with an IFU can be plugged
    anywhere within a 14$\amin$ radius around the IFU.  All harnesses
    have 2 sky ferrules for each V-groove block, with the exception of
    the mini bundle harnesses which have 3 sky fibers terminated into
    their single V-groove block. The bundle is mounted to the anchor
    blocks inside the cartridge between the IFU transition and the
    V-groove transition leaving 0.7~m length free for plugging.
\label{fig:harness-overview}}
\end{figure*}

\subsection{Slit Density}
\label{sec:slit-density}

Choosing the acceptable spacing between V-grooves, ``slit density,''
was a major design decision balancing the total number and size of our
IFU complement with the fidelity of our extracted spectral signal.
Based on (i) an initial analysis of BOSS calibration data and (ii)
measurements of derived properties (e.g., rotation curves) from
simulated data cubes with deferent amounts of crosstalk, this design
trade warranted fabrication of proto-type single-fiber and IFU
hardware for further testing. Data taken using these hardware
were analyzed (1) to confirm
consistency in levels of crosstalk with the BOSS calibration data; (2)
to compare scientific analysis of the same galaxies observed during
the test-run data using IFUs with different slit densities; and (3) to
understand the impact of fiber spacing on spectrophotometric
calibration.

\begin{table}
\caption{\label{tab:crosstalk}Crosstalk as a function of fiber separation}
  \begin{center}
    \begin{tabular}{cccccc}
      \hline
      fibers & Separation & Center & Center & Edge & Edge\\
                & ($\mu$m) & Red & Blue & Red & Blue\\
      \hline \hline
      1000 & 266.0 & 0.006(0.30) & 0.007(0.30) & 0.032(0.45) & 0.002(0.17)\\
      1300 & 204.6 & 0.031(0.20) & 0.035(0.20) & 0.099(0.35) & 0.018(0.13)\\
      1500 & 177.3 & 0.061(0.10) & 0.067(0.10) & 0.158(0.25) & 0.041(0.07)\\
      1700 & 156.5 & 0.091(0.02) & 0.100(0.02) & 0.220(0.10) & 0.075(0.01)\\
      \hline
    \end{tabular}\\Values are in \%; parenthetical values are the fraction of peak flux where
profiles cross.
  \end{center}
\end{table}

Using BOSS sparse plug data (flats taken with only every
4\textsuperscript{th} fiber plugged) we simulated data with different
slit densities ranging from the current BOSS configuration allowing
1000 fibers total to 1700 fibers total corresponding to the physical
limit for a $\sim$150~$\mu$m OD fiber. The results are summarized in
Table \ref{tab:crosstalk}, which gives the crosstalk (in \%) for each
fiber separation in the two spectrograph arms at the center (best
case) and the edge of the field (where aberrations are strong): In the
worst case (1700 fibers, red edge) we expect ~22\% crosstalk (16\% for
1500 fibers). Parenthetical values give the percentage of peak flux
where neighboring profiles cross. In all other cases this study
indicates crosstalk levels below 10\%.  As a comparison, the crosstalk
between neighboring fibers on the sky due to the focal-plane PSF
(i.e., from atmospheric smearing) is 6\% given the median seeing at
APO (the mapping from IFU to the slit ensures that fibers adjacent on
the slit are also adjacent on the sky). In terms of the effect on
MaNGA's science goals, extensive simulations of the measurement of
abundance patterns, abundance gradients, velocity fields and extracted
velocity curves, and velocity dispersion fields show that the majority
of science cases do not suffer from slit crosstalk levels as high as
20\%. The conclusion is that a density corresponding to 1500 fibers
would be acceptable, given experience with other IFU systems, which
generally tolerate and are able to correct for higher levels of
crosstalk.

Based on the above analysis, we designed and fabricated IFU and
single-fiber harnesses with slit spacings varying between 266~$\mu$m,
204~$\mu$m, and 177~$\mu$m. These prototype hardware were otherwise
similar to the final production hardware we describe below. During
test runs in December 2012 and January 2013 using this prototype
hardware, we collected stellar calibration as well as galaxy data. The
stellar data was used to test PSF reconstruction, and at the same time
provide a measure of post-extraction crosstalk between fibers. The
adapted version of the BOSS pipeline does account for and subtract
crosstalk. The remaining level of crosstalk between neighboring fibers
after subtraction is less than 3\%, and $\sim$5\% at the edges of the
field.

Overall, we find the simulations very reliably predict the actual
measured profile in the test data. We conclude that for the galaxy
IFUs a configuration of $\sim$1500 fibers total spaced at 177~$\mu$m
giving raw crosstalk levels of $\sim$20\% in the worst case and
$\sim$10\% over most of the field yields the most efficient survey.
Cross-talk subtraction during data reduction and extraction reduces
the level to less than 5\% (worst case) which is less than the
cross-talk introduced between neighboring fibers in the telescope
focal plane during normal seeing conditions.

To confirm that these cross-talk levels do not impact our scientific
requirements, we observed the same galaxy with bundles of different
slit fiber spacing. One observation used a 127-fiber IFU bundle with a
third each of the fibers mapped into to 260, 204 and 177~$\mu$m
V-groove spacings on the slit. A second observation targeted the same
galaxy with a 127-fiber IFU where all fibers map uniformly to
204~$\mu$m V-groove spacing. Both data sets were processed by the
MaNGA pipeline, capable of handling the different groove
spacings. Stellar and gas kinematics plus a comprehensive set of
spectral line indices were estimated from the extracted spectra and
compared. Differences in estimated velocities, velocity dispersions
and absorption line indices are less than 3\% across the central
regions of the IFU bundle, with a maximum of 5\% at the edges, with no
discernible systematics introduced by the differences in the V-groove
spacings.

However, analysis of stars observed in single fibers and IFUs led us
to conclude that accurate spectrophotometric calibration in the
presence of differential atmospheric diffraction and fiber diameters
comparable to the seeing required observations through 'mini-bundles,'
or 7-fiber IFUs \citep{Yan+2014a}.  In order to accurately reconstruct
the PSF location to determine the flux amplitude at each wavelength it
was found that a fiber spacing of 204~$\mu$m was desirable to further
reduce corrected cross-talk levels to below 1\%.  Consequently, the
final design of our fiber pseudo-slit consists of 177~$\mu$m spacing
for IFUs and their associated sky fibers, and 204~$\mu$m spacing for
the calibration bundles and their associated sky fibers. In addition,
gaps of 246~$\mu$m were placed between fiber groups arranged in blocks
containing between 21 and 39 fibers (see \ref{sec:v-groove-block}) to
permit measurements of the wings of the PSF and scattered light. The
block separation was determined based on experience with the BOSS
fiber system to allow for 624~$\mu$m between fiber edges at the ends
of two adjacent blocks. The net effect of these considerations
resulted in slightly less than 1500 fibers per cartridge in our final
design.

\section{IFU and Harness Design}
\label{sec:ifu-design}

A fiber harness is a grouping of fibers and their associated mounting
hardware that transfer the light from the telescope focal surface to
the BOSS spectrographs. We show an overview of the components of a
MaNGA fiber harness in Fig.~\ref{fig:harness-overview}, which measure
$\sim$1.7~m in length. Depending on the SDSS survey configuration using the BOSS
spectrographs, 31 to 50 harnesses are included in each cartridge for
each survey, with each harness containing between 20 and 135
fibers. For MaNGA cartridges (shared with the APOGEE-2 survey) there
are 21 MaNGA harnesses feeding the BOSS spectrographs and 10
single-fiber APOGEE harnesses feeding the APOGEE spectrograph.

In addition to the fibers, a harness consists of cabling and
anchor points to protect and attach them within a cartridge, fiber
ferrule terminations for plugging into plates at the telescope focal
plane, and fiber V-groove termination blocks which are grouped
together to form pseudo-slits feed the spectrograph collimators.

MaNGA terminates 1423 fibers into 21 harnesses of six different types.
The larger number of fibers are accommodated with a tighter spacing
within the V-grooves that populate the spectrographs' pseudo-slits as
discussed in the previous section. The six harness types contain
between 21 and 135 fibers, depending on the IFUs they contain. The
fibers are terminated into V-groove blocks (between one and four,
depending on the total number of fibers). The V-groove blocks (see
\ref{sec:v-groove-block}) come in different sizes to accommodate the
different harness fiber counts so that harnesses do not share V-groove
blocks. Five of the six harness types contain single IFUs intended for
galaxy targets, with 19, 37, 61, 91, or 127 fibers plus associated sky
fibers (2, 2, 4, 6, and 8 respectively). The sixth harness type
contains three mini-IFUs (``mini-bundles'') of seven fibers each plus
three associated sky fibers; these are used for observing standard
stars for spectrophotometric calibration. MaNGA terminates each IFU
and each sky fiber into its own, separately pluggable ferrule. The
total number of pluggable ferrules is 121, consisting of 17 galaxy
IFUs, 12 calibration mini-bundles, and 92 sky fibers. These map to 44
V-groove blocks split between two spectrographs.

Each harness is capable of being plugged into any location on the
focal plane. The sky fibers associated with the harness may plug
within a 14$\amin$\ radius of their associated IFU.  This is done to
keep the plugging tractable and to maintain locality of sky
sampling. Based on our analysis of extant BOSS data and results
gathered from engineerings runs using proto-type MaNGA hardware, we
determined that sky-subtraction \citep{Law+2014} requires sampling
over the full slit, as well as local sampling on the sky and on each
V-groove block. Two sky-fibers are placed at the edge of the V-groove
block. In the calibration mini-bundles a third sky-fiber is placed one
groove in from the end.  This third sky fiber is then adjacent to one
sky fiber and one IFU fiber.

Walking through the components of a fiber harness from left to right
in Fig.~\ref{fig:harness-overview}, the ferrules are the part of the
harness that engages with the plug plate and contains the polished
fiber input surface. The central ferrule contains the IFU fibers
packed into a hexagonal array. Sky ferrules each contain a single
fiber terminated in their center.

The IFU transition is 280~mm back from the front of the
ferrule where the sky fibers and IFU fibers meet to transition into a
single fiber bundle.  The length of this transition was determined to
be necessary for ensuring that sky fibers could be plugged around the
IFU and not fall out during operations. Shorter lengths led to
plugging difficulties and dropped fibers in lab tests. Longer lengths
are workable but become unruly during plugging operations. The
transition length sets the 14$\amin$\ patrol radius for sky fibers around
their IFU.

The V-groove block transition is located 860~mm back from the IFU
face. This is where the fibers break into their V-groove block groups
and enter a smaller cross-section jacket for the transition into the
slit plate. The slit-plate transition occurs roughly $\sim$260 mm from
the V-groove termination. The detailed routing (described below)
ensure that the bare fibers are well protected and mechanically
stable.  Smaller harnesses, such as those for the 19 and 37 fiber IFUs
as well as the mini-bundles, only contain one V-groove block, so this
transition simply acts to reduce the size of the fiber jacketing. The
large IFUs contain from two to four V-groove block branches.  V-groove
blocks contain between 21 and 39 fibers, with the maximum number
limited by the fiber spacing and the overall size of the block, which
we limit to under 8~mm. The latter restriction keeps defocus at the
spectrograph input focal surface below 2~$\mu$m for an f/4 beam, given
the flat polished exit surface of the V-groove blocks and the
spherical collimator 1264 mm radius of curvature.  With the exception
of the mini-bundles where there are three bundles (and 3 sky fibers)
on a single V-groove block, no IFUs share V-groove blocks.

\subsection{Fiber Selection}
\label{fiber-selection}

\begin{figure}[tbh]
  \centering
  \includegraphics[width=8cm]{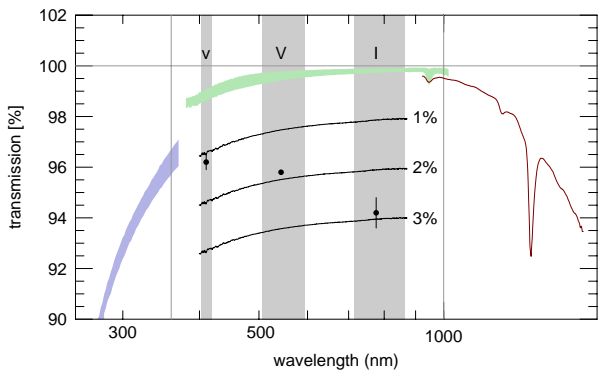}
  \caption{Fiber transmission for production-build Polymicro FBP
    step-index multi-mode fiber scanned at 2~m length by Polymicro
    (green and purple polygons). The 2\% discrepancy between 380 and
    500~nm corresponds to the transition region in the Polymicro UV
    (purple) and VIS (green) spectrometers; polygon thickness
    represents variation between different scans.  We expect the green
    curves to be representative of true performance $>$ 380 nm; and
    the purple curves to representative $<$ 380 nm. A long-wavelength
    scan of our prototype fiber draw (red curve; same fiber type and
    core:clad:buffer sizes) shows attenuation increase in the NIR, but
    outside our band-pass. Model transmission curves for 1, 2, and 3\%
    Fresnel loss per surface are shown in black and compared to
    measurements made for AR-coated prototype fibers in the
    band-passes given by grey regions (Stromgren v plus Johnson V and
    I). Measurements are within $\pm$1\% of
    expectations.\label{fig:FBP-transmission}}
\end{figure}

The only vendor with proven, high-performance fiber suitable for
astronomical application in the visible and NIR is Polymicro. Their
best product is FBP, a step-index fused-silica fiber with 0.22~NA, low
attenuation between 275 and 2100~nm, low focal-ratio degradation
(FRD), and polyimide buffer. Estimates of fiber attenuation provided
by the vendor, corresponding to the 1.7m length of the MaNGA
harnesses, are compared in Fig.~\ref{fig:FBP-transmission} with our
laboratory measurements of as-built harness transmission.

\begin{figure*}[tbh]
  \centering
  \includegraphics[width=0.9\textwidth]{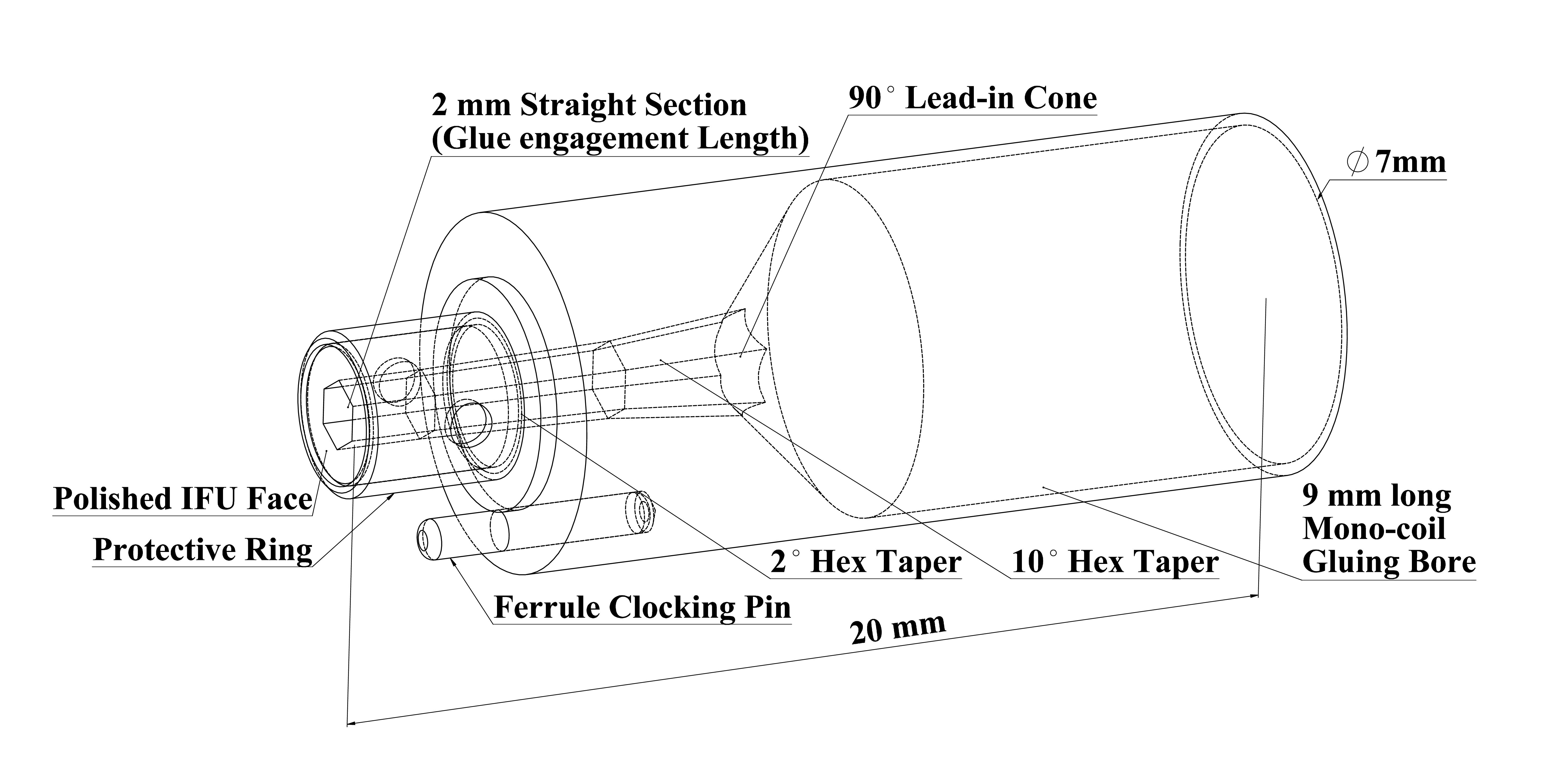}
  \caption{One of six kinds of hexagonal aperture IFU ferrules, shown
    in cross-section. Different ferrule types vary only by the
    clearance allowed at the hexagonal taper and straight section. The
    fibers in the IFU face are glued, polished and anti-reflection
    coated before the protective ring is added to the front of the
    ferrule.\label{fig:ferrule-cross}}
\end{figure*}

Existing BOSS fiber cartridges use Polymicro FBP 120:170:190
(core:clad:buffer diameter) fiber. The active core is 2$\asec$\ on
sky.  The median seeing at Apache Point Observatory is 1.1$\asec$,
70~$\mu$m at the focal plane, so larger core sizes than 120~$\mu$m
significantly under-sample the PSF. The spectrograph also is designed
to be critically sampled for this fiber-core diameter, delivering a
monochromatic FWHM at the detector of 45~$\mu$m\ $\sim$3 pixels.  The
camera-collimator focal ratios demagnify the fiber image by a factor
of 2.67, and between 18 and 30~$\mu$m of the monochromatic FWHM is
contributed (in quadrature) by optical aberrations
\citep{Smee+2013}. Larger fiber sizes would degrade spectral
resolution; smaller fibers would be under-sampled and only marginally
improve the monochromatic FWHM while collecting less light. These
considerations led us to retain the BOSS fiber core size of 2$\asec$.

However, the BOSS fibers have thick cladding and buffer.  Because
optimizing the packing density is a design goal, we worked with
Polymicro to design a fiber with 120~$\mu$m core that minimized
cladding and buffer thickness while guaranteeing, respectively, (1)
total internal reflection in our band-pass and (2) adequate protection
in laboratory handling during assembly and routine use on the
telescope. We identified an existing form to draw 120:132:150~$\mu$m
diameter core:clad:buffer fiber using the broad-spectrum FBP glass
recipe and polyimide buffer. Identifying and using an existing form
made the product affordable.

We have adopted this FBP 120:132:150 product for both prototyping and
production. Polymicro demonstrated this FBP fiber product yielded the
same low-attenuation within 350-1000~nm as their standard product
(Fig.~\ref{fig:FBP-transmission}).  We subsequently verified its high
throughput and excellent FRD properties in the lab and on sky. The
live-core fill factor reached by this fiber in our ferrules is
$\sim$56\%, as discussed below.

With this custom form, Polymicro committed to delivering an outside
diameter (OD) tolerance of 150~$\mu$m~$\pm$5~$\mu$m.  We requested and
received verification of this by way of the draw tower measurements,
which take final outside diameter measurements at a cadence of 1
measurement/m as the fiber is drawn. We found that while the final
diameter of the fiber will fall within a band of $\pm$5~$\mu$m\ for a
given draw the diameter is stable at the 1~$\mu$m\ level.  Our
specific draw of fiber has an outside diameter of
151~$\mu$m~$\pm$0.5~$\mu$m\ for the full 29~km draw length.  This was a
critical discovery that allowed for significant improvement in packing
regularity discussed further in \S~\ref{sec:packing-regularity} below.

\subsection{Ferrule Design}
\label{ferrule-design}

\subsubsection{IFU Shape}
\label{sec:ifu-shape}

We have chosen a hexagonal IFU shape as a natural cylinder-packing
geometry that also approximates the characteristic shape of galaxies
on the sky.  While there is a wide range of apparent ellipticity in
galaxy light distributions, the mean ellipticity is such that sampling
with a near-circular geometry places a relatively small fraction
(~30\%) of fibers off-source. Random variations in ellipticity
distributions, field-to-field, make elongated bundle sizes intractable
for efficient use of fibers in targeting, particularly because
position-angle variations cannot be accommodated without risking
fiber-stress, ensuing FRD and light loss. More importantly, in cases
where sources are very elongated, i.e., highly-inclined disks, early
results from SAMI (Sydney-Australian-Astronomical-Observatory
Multi-object Integral-Field Spectrograph; \citealp{Croom+2012}) have
shown potential for detection of gas outflows from star-forming disks,
one of MaNGA's science objectives. For all of these reasons, we expect
our choice of densely packed, hexagonal fiber IFUs will serve MaNGA
well. This choice of packing discretizes the number of possible fibers
per IFU,

\begin{displaymath}
  N_{\rm fiber} = 1 + \sum_{n=1}^{N_{\rm ring}} 6n = 1 + 3N_{\rm ring}(N_{\rm ring}+1)
\end{displaymath}

\noindent where $N_{\rm ring}$ is the number of hexagonal rings or
layers beyond the central fiber.  The first several values for $N_{\rm
  fiber}$ for $N_{\rm ring}$ = 1 .. 8 are 7, 19, 37, 61, 91, 127, 169,
217.

\subsubsection{EDM-Enabled Tapered Hexagonal Bore}
\label{sec:edm-bore}

The primary method for building bare-fiber IFUs has been a
mold-release process pioneered for DensePak on the WIYN 3.5m telescope
\citep{Barden+1998}, and subsequently adopted for SparsePak
\citep{Bershady+2004} and PPak \citep{Kelz+2006} on the WIYN and Calar
Alto 3.5m telescopes, respectively. With this method fibers are
individually placed into a mold of a known shape and potted with epoxy
before polishing. This method yields IFUs of exceptional fiber
regularity but suffers from two drawbacks. First, there is often a
significant lower performance in the edge fibers that are in contact
with the mold \citep{Bershady+2004}. This can be mitigated (e.g.,
\citealp{Kelz+2006}) by including sacrificial edge fibers, acting as
stress buffers for the live fibers in the interior of the IFU.
Another alternative is to pack fibers in an array of micro-tubes
\citep{Arribas+1991}. This also mitigates mold-induced stress but
decreases packing density as well. More significantly, these
approaches to building IFUs are very labor intensive. Given shear quantities of IFUs to be
constructed for MaNGA the mold or tube-form methods were considered a secondary
option.

A newer method of building densely packed bundles was developed by
\citet{Bland-Hawthorn+2010} for use on the AAO spectrograph.  Here,
fibers are stripped of their polyimide buffer, and the cladding is
etched down to a minimal size. The fiber bundles are then slid into a
glass tube of circular cross-section, which in turn slides into a
aluminum ferrule for protection.  The entire assembly is then potted
with a soft silicone glue which fills the interstitial area between
fibers.  These so-called ``Hexabundles'' derive their name from a
version where the fibers are lightly fused at one end such that their
active cores take on a hexagonal shape. This process is able to
achieve very high fill ratios of 84\%, although not without some costs
in both FRD and robustness of the design
\citep{Croom+2012,Bryant+2014}. The process for construction is also
rather involved with wet chemical work needed for stripping the
cladding and etching the buffer. While the packing density is very
high, the packing ends up being irregular, which we find in
simulations \citep{Law+2014a} is a detriment for image reconstruction.

Using the Hexabundle fabrication method as a template, we attempted to
improve the concept in terms of fiber packing regularity and
performance, albeit at the sacrifice of packing density, through a
modification of the manufacturing technique.  Principally we changed
the packing form from a circular to a hexagonal cross-section to
improve packing, and retained the (minimal) fiber buffer to improve
performance. The trick in the procedure proved to be integrating the
fibers into a single ferrule design where the fibers themselves are
pushed into a stainless steel sleeve and glued in place.  This was
possible through use of a wire electro-magnetic discharge (EDM)
machine at U.\ Washington capable of handling EDM wire thickness as
small as 100~$\mu$m. With this thickness wire we were also able to
produce stainless steel ferrules of arbitrary internal size and shape
for our 151~$\mu$m OD fibers. Hexagonal cross-sections proved to be
the ideal shape to produce a well ordered hexagonal packing inside the
IFU.

The key aspect of our design is to facilitate the natural packing and
ordering of the fibers within the ferrule by implementing a taper that
serves as a funnel for the fibers. We started with a simple 45 degree
cone at the back of the ferrule, funneling the fibers towards a set of
progressively narrower hexagonal tapers finally ending in a 2~mm
section of strait hexagonal bore which acts as the gluing interface
for the fibers (Fig.~\ref{fig:ferrule-cross}).  This novel tapered
design was proven in the lab at U.\ Washington and U.\ Wisconsin as
well as by our assembly vendor (C~Technologies, referred to as C-Tech)
to provide easy and reliable fiber insertion with extremely tight
tolerances. One item worth noting is that the application of alcohol
into the ferrule during insertion served as an excellent lubricant and
provided additional surface tension to naturally form the fibers into
a hexagonal pattern. As we discuss below, we have achieved better than
3~$\mu$m RMS in relative location of the fibers from an ideal packing
while not inducing {\it any} measurable FRD above and beyond what is
observed in single-fiber ferrules (see
\S~\ref{sec:laboratory-measurements-of-ifus}).

The ferrules themselves can be built to arbitrary size with only minor
changes to the coding of the EDM machine, enabling us to produce our
6 ferrule sizes with one set of tooling and several variations of
manufacturing code.  The final ferrules can be made for under \$300
USD each to 3~$\mu$m precision at a near 100\% pass rate through
quality assurance.  The assembly of the ferrule end of the IFU then
takes merely a few minutes (followed by curing and subsequent
polishing).

\begin{figure}[tbh]
  \centering
  \includegraphics[width=8cm]{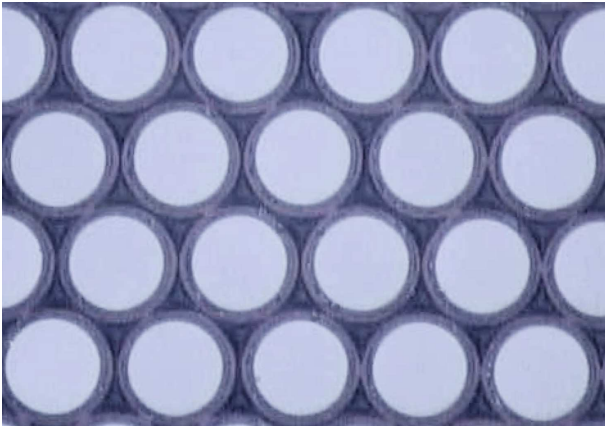}
  \caption{Close-up of a section of a 127-fiber IFU showing the
    packing regularity with a median fiber spacing 151.18~$\mu$m and
    median positional error of 3.5~$\mu$m.\label{fig:fiber-closeup}}
\end{figure}

\subsubsection{Packing Regularity}
\label{sec:packing-regularity}

\begin{figure*}[tbh]
  \centering
  \includegraphics[width=0.45\textwidth]{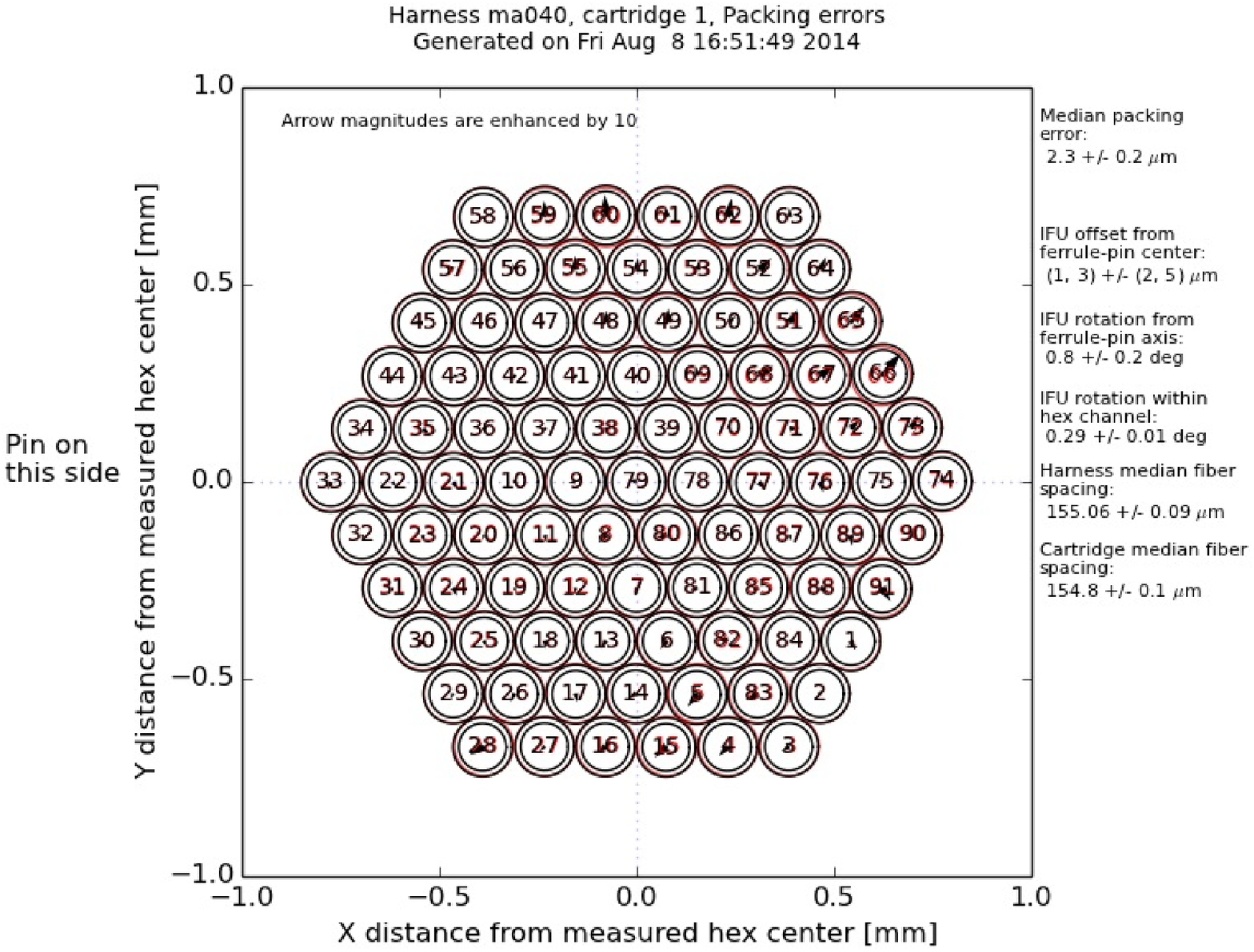}
  \includegraphics[width=0.45\textwidth]{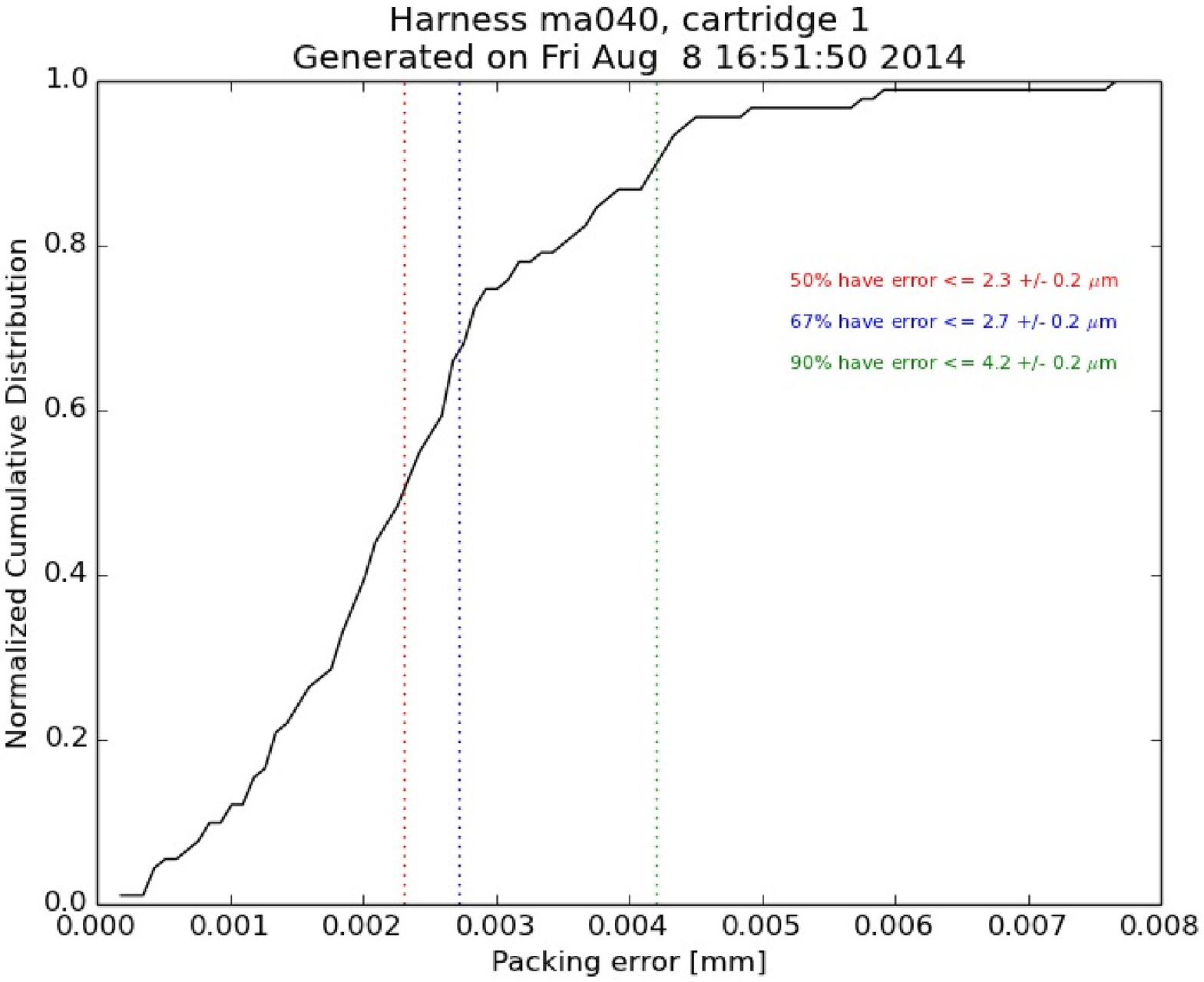}
  \caption{IFU metrology as generated for every IFU harness from the
    SDSS test-stands for a 91-fiber IFU (MA.040). Positional error
    maps for all fibers (left) provide vectors showing the direction
    and magnitude of the fiber position relative to its intended {\it
      relative} position within an ideal fiber hexagon (arrows are
    amplified ten times). Absolute positional errors (translation and
    rotation) of the fiber hexagon with respect to the center of the
    outer ferrule diameter and a locating pin are listed on the side
    of the map, along with the mean fiber spacing, IFU identifier and
    time of measurement. The normalized, cumulative histogram (right)
    of fiber relative positional errors shows that 67\% of the fibers
    have positional accuracies better than 3~$\mu$m and 90\% are
    better than 5~$\mu$m.  All information is tabulated and stored in
    an on-line database connected to the data reduction pipeline and
    data archive.\label{fig:positional-errors}}
\end{figure*}

\begin{table}
\caption{\label{instrument:tab:ferrule_tolerance} Production ferrule tolerances}
  \begin{center}
    \begin{tabular}{cccc}
      \hline
\multicolumn{2}{c}{\# IFU fibers} & Clearance$^\dagger$ & Hex Size \\ \cline{1-2}
total & diagonal & (mm) & (mm) \\
      \hline
      \hline
        7 &  3 & 0.003 & 0.413 \\
       19 &  5 & 0.005 & 0.675 \\
       37 &  7 & 0.007 & 0.937 \\
       61 &  9 & 0.009 & 1.199 \\
       91 & 11 & 0.011 & 1.461 \\
      127 & 13 & 0.013 & 1.723 \\
      \hline
    \end{tabular}\\
    $^\dagger$ The clearance is the amount by which the ferrule
    hexagon is oversized relative to the nominal size taken up by the
    fibers assuming perfect packing.
  \end{center}
\end{table}

\begin{table}
\caption{\label{instrument:tab:first-article-packing-regularity} First Article Packing Regularity }
  \begin{center}
    \begin{tabular}{cccc}
      \hline
IFU Size & \multicolumn{2}{c}{Packing Error ($\mu$m)} \\ \cline{2-3}
      (\# fibers) & 50\% & 90\% \\
      \hline
      \hline
      7 & 1.1 & 1.8\\
      19 & 2.0 & 3.8\\
      37 & 1.5 & 2.7\\
      61 & 3.6 & 6.3\\
      91 & 2.7 & 5.1 \\
      127 & 4.2 & 7.3\\
      \hline
    \end{tabular}
  \end{center}
\end{table}

Simulations of cube reconstruction of MaNGA-like observations
suggested that a regularity requirement of 5~$\mu$m RMS fiber
positional error would be needed to adequately reconstruct images
within data cubes and to keep degradation in the delivered FWHM and
axial ratio of a point-like source below 10\% \citep{Law+2014a}

During early development, sets of 19 and 127 fiber ferrules with
different IDs from 5~$\mu$m to 25~$\mu$m of wall clearance were built
to conduct packing tests (see Fig.~\ref{fig:fiber-closeup} for an
image of one of our proof-of-concept early IFUs). IFU ferrules with 19
fibers and wall clearances down to 5~$\mu$m packed with ease and
showed a much higher level of regularity than expected. Similarly the
smallest 127-fiber ferrule with a wall clearance of 10~$\mu$m packed
with little resistance.  A key feature in our ability to achieve such
small clearances inside the IFU was the prior knowledge of the fiber
diameter to within 1~$\mu$m.  This allowed us to dramatically reduced
the needed clearance by custom building the ferrules to match the
fiber draw on hand.  Another key feature, which allowed for very tight
manufacturing tolerances was the use of EDM manufacturing methods
discussed above that allowed for control of the ferrule internal
clearance to within 3~$\mu$m.

Manufacturing of prototype IFUs showed that fibers have a tendency to
group themselves together inside the slightly over-sized hexagonal
aperture in a tight array that is offset to one side of the ferrule,
thereby preserving better than expected packing density and regularity
(see Fig.~\ref{fig:fiber-closeup}). Prototype ferrules which had a wall
clearance of 6~$\mu$m for the 19 fiber ferrules and 15~$\mu$m for the
127 fiber ferrules showed extremely good fiber regularity of between 2
and 4~$\mu$m RMS. The global position offset of the pattern was larger
at around 20~$\mu$m for the 127 fiber IFU. This global offset can be
attributed to the fiber biasing to one side of the ferrules and to the
offset of the inscribed hexagon in the OD of the ferrule. This offset
is measured as part of our metrology quality assessment described
below, and is accounted for in the data reduction pipeline \citep{Law+2014}.

Table~\ref{instrument:tab:ferrule_tolerance} gives the final ferrule ID
tolerances of the production ferrules.  These values yield exceptional
packing regularity while being easy to assemble. The exact value used
for the ferrules is adapted to match the exact OD of the fiber used in
production based on measurement and documentation of the fiber OD
during the fiber draw.

\begin{figure*}[tbh]
  \centering
  \includegraphics[width=0.9\textwidth]{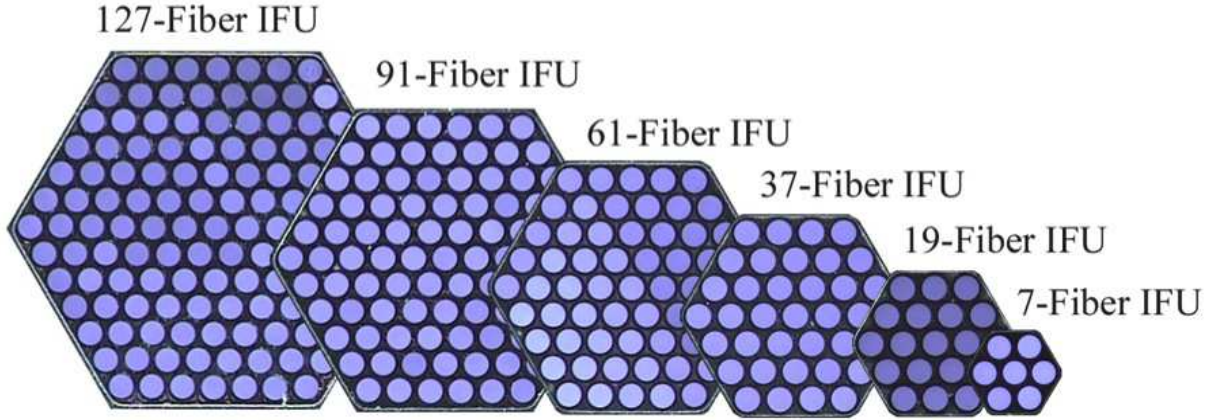}
  \caption{Images of the fibers in MaNGA IFUs ranging from 7 to 127
    fibers (right to left).\label{fig:ifu-closeups}}
\end{figure*}

Table~\ref{instrument:tab:first-article-packing-regularity} shows the
packing error associated with the first cartridge build completed in
January of 2014, and Fig.~\ref{fig:positional-errors} shows the
metrology data for a typical IFU. Not surprisingly, there is a slight
trend towards higher positional error in the packing regularity as the
bundle size increases. This is due to the larger assembly clearances,
although even the largest size bundle perform better than the goal of
5~$\mu$m RMS fiber positional accuracy. Finally,
Fig.~\ref{fig:ifu-closeups} shows closeup images of our IFUs.

\subsubsection{Fill Factor}
\label{sec:fill-factor}

The size and fill-factor of our IFUs are optimized for our target
selection \citep{Wake+2014} that enables our science goals of
achieving coverage out to 1.5 to 2.5 \Reff, sufficient S/N to measure
stellar composition gradients and kinematics, and the acquisition of a
large sample of 10,000 galaxies.  For example, more sparsely spaced
fibers can be grouped into wider-field IFUs to sample galaxies with
larger apparent size, at the sacrifice of either radial coverage or
S/N per pointing on the galaxy. More widely spaced fibers also make
PSF characterization (and therefore data-cube reconstruction) more
difficult.  For these reasons, well-informed by simulations, we
concluded that the most desirable IFU configuration was one where the
fibers were as closely spaced as possible \citep{Law+2014a} Concerns
over durability (fibers are physically handled daily in operations)
and throughput uniformity eventually lead us to leave the buffer
intact.  Even with the intact buffer, our IFUs reach a fill factor of
56\%.

\begin{table}
\caption{\label{instrument:tab:cartridge-complement} Cartridge Complement }
  \begin{center}
    \begin{tabular}{ccccccc}
      \hline
\multicolumn{4}{c}{IFU Size} && Sky fibers & IFUs \\ \cline{1-4}
fibers & rings & corner ($\asec$) & flat ($\asec$) && per IFU & per cart\\
      \hline
      \hline
        7 & 1 &    7 & 6 && 1 & 12 \\
       19 & 2 & 12 & 10.4 && 2 & 2 \\
       37 & 3 & 17 & 14.7 && 2 & 4\\
       61 & 4 & 22 & 19.0 && 4 & 4\\
       91 & 5 & 27 & 23.3 && 6 & 2\\
      127 & 6 & 32 & 27.7 && 8 & 5\\
      \hline
    \end{tabular}
  \end{center}
\end{table}

\subsubsection{Size Distribution}
\label{sec:size-distribution}

The IFU complement is optimized in its size distribution given our
science goals, available slit real estate, and the natural apparent
size and redshift distributions of target galaxies.

\begin{figure}[tbh]
  \centering
  \includegraphics[width=0.45\textwidth]{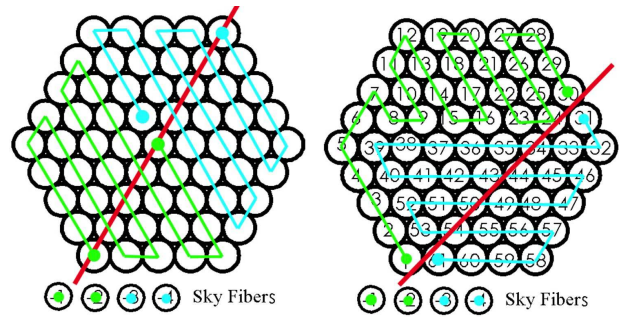}
  \caption{Mapping of the 61-fiber IFU to the spectrograph slit (left)
    using an initial serpentine pattern and (right) with the final
    optimized pattern that minimizes cross-talk between galaxy signal
    and sky fibers. Fibers with dots indicate the start and end of a
    single V-groove block. Next to each of these dotted fibers at the
    edge of the V-groove block is a sky fiber. Spacing between fibers
    in V-groove blocks is 177~$\mu$m for all IFUs with 19 or more
    fibers, while spacing between the edge fibers of adjacent V-groove
    blocks is 624~$\mu$m. Black diagonal lines indicate how a
    highly-inclined disk galaxy would need to be positioned and
    oriented to significantly contaminate more than one sky
    fiber.\label{fig:map-old-new}}
\end{figure}

\begin{figure*}[tbh]
  \centering
  \includegraphics[width=0.6\textwidth]{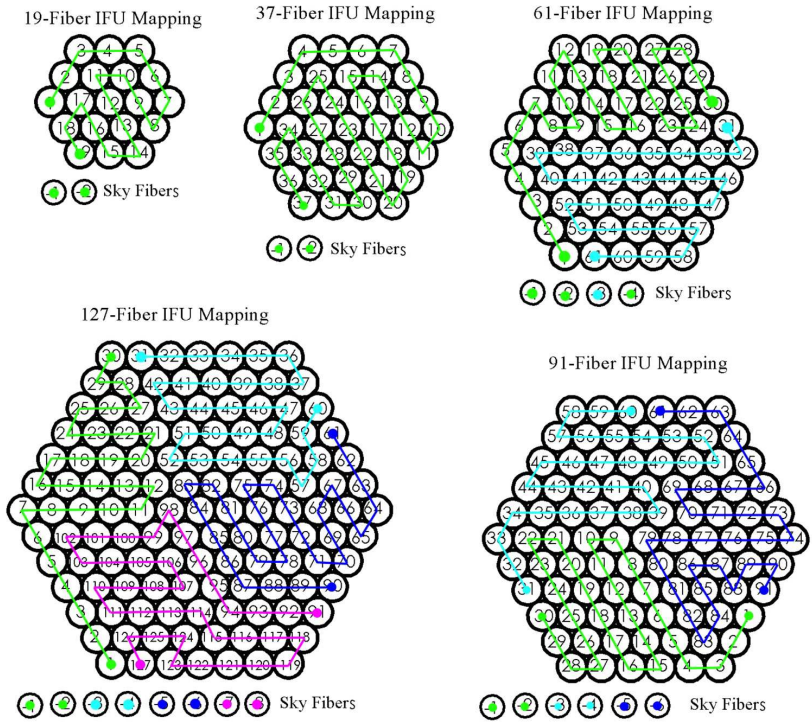}
  \caption{Mappings of fibers from the IFU in the telescope focal
    surface to the spectrograph slit for all IFU sizes (mapping for
    61-fiber IFU is shown in previous figure). Different colors trace
    fibers mapped into different individual V-groove blocks.  Fibers
    with dots in them are the fibers adjacent to the sky fibers at
    either end of the V-groove block.\label{fig:ifu-maps}}
\end{figure*}

An iterative process of sample design was then conducted to determine
the most optimal distribution of the IFU sizes to match with the above
mentioned goals.  An upper limit of 127 fibers/IFU was implemented as
the largest size IFU available primarily because larger IFUs would be
costly in terms of total sample size for the survey.  The optimal
distribution was then based on best matching of the available sized
IFUs 19, 37, 61, 91, and 127 to the 1.5~\Reff\ coverage minimum for 7
degree\textsuperscript{2} fields tiled over the Sloan imaging data
base. A closeup image of an IFU of each size is shown in
Fig.~\ref{fig:ifu-closeups}.  A full description of this optimization
process is presented in \citet{Wake+2014}. The results of sample
selection lead to a single-cartridge complement summarized in
Table~\ref{instrument:tab:cartridge-complement}.  Angular sizes of
the IFUs are measured as edge-to-edge distances of the active cores
assuming a plate scale of 60~$\mu$m per arcsec.  The total fiber count
is 1423. Of these, 92 are dedicated to sky, 84 are used for
spectrophotometric calibration in 12 seven-fiber IFUs, and 1247 fibers
are spread between 17 IFUs.

To populate the slit optimally, we set the gap between the edge fibers
of neighboring v-groove blocks to 624~$\mu$m (see
\ref{sec:slit-density}) to allow for characterization of the wings of
the fiber light profile as well as for obtaining a handle on scattered
light across the CCD. This sets the distance between the V-groove
blocks themselves to 313~$\mu$m. We also limited how far out on the
slit we were willing to go to keep aberrations and hence cross talk to
the limits determined above and limit the amount of vignetting. We
then ran optimizations of the slit area to determine the distribution
of IFU fibers between both slits to match best with the optimal
complement to cover our galaxy sample determined by \citet{Wake+2014}.

\subsubsection{Fiber Mapping}
\label{sec:fiber-mapping}

Several considerations made a deterministic mapping of fibers between
the 2-dimensional distribution in the IFU and the 1-dimensional
distribution in the V-groove blocks along the spectrograph slit a
critical aspect of the design. To localize crosstalk at the telescope
focal surface at the spectrograph input, we required fibers adjacent
on the slit were also adjacent in the IFU.
To avoid potential systematic errors along the slit or within the bundle
to be mistaken for properties of the observed source, we eliminated
spiral mapping patterns.

A more subtle consideration was the effect of the galaxy light on the
sky fibers. In the telescope focal plane, sky fibers are free to be
plugged within a 14$\amin$ radius of their associated IFU.  At the
spectrograph input, the sky fibers associated with each IFU are placed
on either end of the IFUs V-groove blocks.  As noted above, this
enables us to sample sky both local to the IFU input on sky and in the
spectrograph. The effect of an estimated 20\% (worst case) cross-talk
between fibers means source signal from the one adjacent galaxy fiber
could be significant.  To keep the sky signal from being overwhelmed
by bright fibers (e.g., the centers of some galaxies), we designed a
mapping that always places source fibers from the outer most ring of
the IFU where the target galaxy surface-brightness will tend to be
lowest at the end V-groove blocks adjacent to the sky fibers.  We
further considered that highly-inclined disk galaxies can have high
signal even out to 1.5~\Reff, thereby potentially degrading some of
the sky fibers due to cross talk. In order to mitigate this risk we
implemented a mapping that, while appearing somewhat chaotic, ensures
no two sky fibers paired in a single V-groove block are in line with
each other across the central axis of the IFU
(Fig.~\ref{fig:map-old-new}).  This allows any single V-groove block
to have at least one sky fiber unaffected by bright areas of the
galaxy--at least for axisymmetric systems--while still meeting all of
our other mapping requirements. Figure~\ref{fig:ifu-maps} shows the
mapping between IFUs and their V-groove blocks chosen to satisfy these
considerations.

\begin{figure}[tb]
  \centering
  \includegraphics[width=8cm]{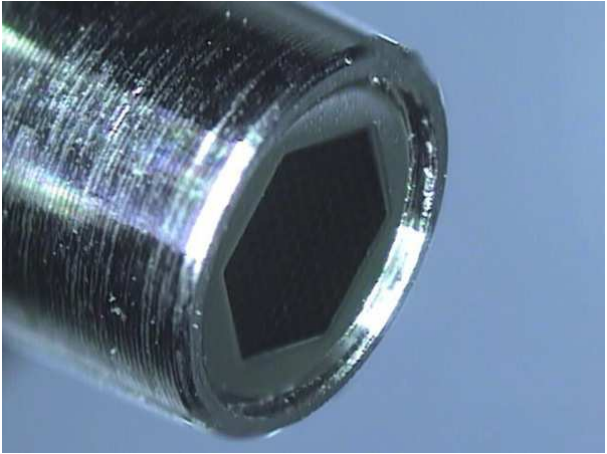}
  \caption{Ferrule protective rings stand 100~$\mu$m above the
    polished face of the IFU, as show in this image.
    \label{fig:ferrule-ring}}
\end{figure}

\begin{figure*}[tb]
  \centering
  \includegraphics[width=0.8\textwidth]{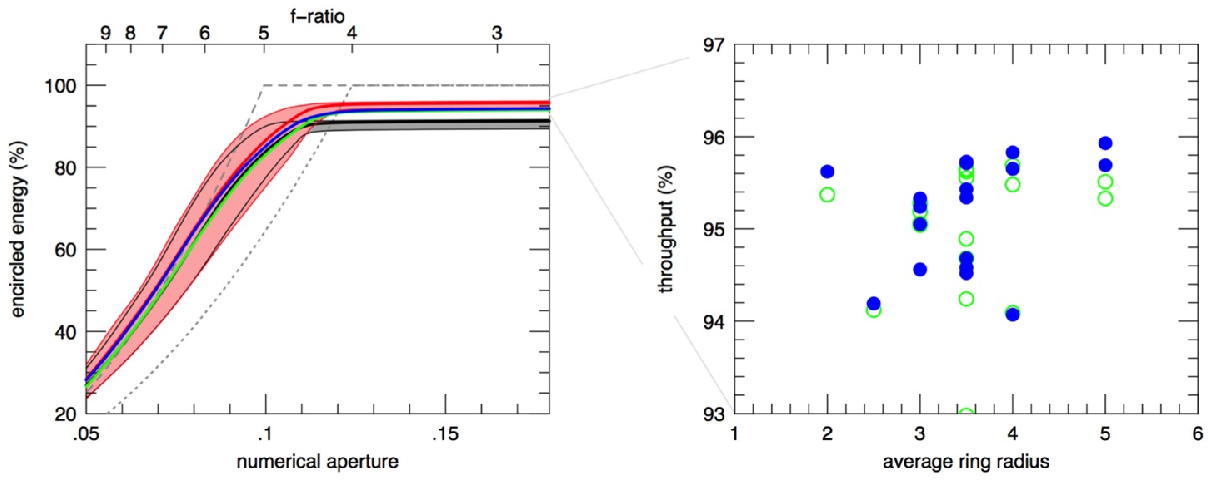}
  \caption{Far-field encircled energy as a function of numerical
    aperture (left) and throughput measurements within a numerical
    aperture of 0.124 (f/4, right) for a test-termination cable with
    127-fiber IFUs on both ends of a 1m length of jacketed fiber.  All
    measurements are made in the Johnson V band, and are color-coded by
    the state of the cable, in order from the original polish (black),
    AR-coated (red), lab life-time tested (green), and finally
    ring-capped (blue).  Grey dashed and dotted curves (left) show a
    uniformly filled f/5 beam (mimicking the fiber input from the
    telescope) and a uniformly filled f/4 beam, corresponding to the
    speed of the spectrograph collimators.  Substantial throughput
    gains are made with AR coating. These gains are unchanged by
    ring-capping within our 0.5\% measurement uncertainty, and do not
    depend on the ring-radius within the IFU.
\label{fig:ring-throughput}}
\end{figure*}

\subsubsection{Protective Rings}
\label{sec:protective-rings}

Each IFU and sky fiber is handled on a daily basis by the plugging
staff at APO.  Their task is to plug the IFUs into the field plates
during the day for use at night.  Plugging technicians plug up to 9
cartridges per day with 421 individual ferrules for MaNGA-APOGEE
cartridges; or up to 1000 individual ferrules for an eBOSS cartridge.
Due to volume, plugging must be quick and efficient.  It is expected
that the face of the IFU ferrule will come into incidental contact
with the aluminum plug plate during normal operations.  IFU fibers
take up a significantly larger area of the ferrule face and become
more susceptible to damage during plugging operations than the single
fiber ferrules used in past SDSS projects.  In lab simulations of
plugging, damage could be seen on the fiber faces after $\sim$150
plugging cycles, half the expected plugging cycles during the survey
lifetime for a single IFU. By 150 plugging cycles damage consisted
primarily of aluminum chips bonding to the surface of the fibers.
Full failure of the AR coating and scratching of the fiber faces did
not occur until 550 cycles, almost two times the expected plugging
cycles for the survey.

Field testing showed significantly more damage to the fiber faces than
from lab tests, however, becoming significant after just 20 plugging
cycles.  Lab conditions were significantly more gentle than mountain
conditions, likely due to the presence of abrasive dust and sand in
the air at APO. This prompted us to design a mechanical buffer to
protect the optical surface of the ferrule.  Because this optical
surface must be be free and clear of mechanical obstructions during
polishing and coating, we settled on a protective ring, pressing onto
the outside diameter of the polished, AR-coated ferrule, and standing
100~$\mu$m above the polished surface. The ring acts as a bumper to
prevent contact of the fiber faces with the plug plate during plugging
operations.  Fig.~\ref{fig:ferrule-ring} shows a closeup image of a
ferrule with protective ring.  This addition becomes the last step
before final performance verification and the fiber assemblies are
integrated into the cartridge.

The ferrule ring is a thin section of 303 stainless steel custom
machined tubing that presses onto the ferrule with a 9~$\mu$m
interference fit. The ring is built to be significantly thinner than
the IFU ferrule wall to insure the majority of the deflection in the
press fit goes into the ring and not the IFU ferrule where it could
cause FRD. Finite element analysis (FEA) was conducted on the 127
fiber IFU ferrule, which has the thinnest wall, to determine the
deflection at the ferrule inner diameter (ID). At the tolerance
extremes of the components, the displacement at the ID boundary is
less then 2.5~$\mu$m. The principal stresses at the ID of the ferrule
were low, generally less than 15~KSI; the ratio of stress in the ring
relative to stress in the ferrule ID was seen to be 4:1 indicating
that the ring was absorbing the majority of the press-fit deformation.

Lab tests confirmed that the ring-pressing does not change the
throughput or FRD properties of either 19 or 127-fiber ferrules
(Fig.~\ref{fig:ring-throughput}).  Measurements of the fiber output
far-field light profile (shape and amplitude) were made before and
after the addition of the press fit rings using equipment described in
\S~\ref{sec:laboratory-measurements-of-ifus}. The results showed no
significant difference after the addition of the press-fit rings.

Further lab life-time testing also confirmed that the protective rings
eliminated damage to the surface of the IFU after more than 500
plugging cycles. Despite prior differences between lab and field
testing, given that damages seem to come primarily from the
interaction of the face of the IFU with the surface of the plug plate,
we anticipate the protective rings will reduce damage in the field as
well. The rings can be removed easily should the IFUs unexpectedly
require re-polishing after a period of service.

\subsubsection{Plug-Plate Interface}
\label{sec:plug-plage-interface}

Fiber location in the telescope focal surface must be accurate and
repeatable in position (to 0.2"), rotation (3$\deg$), and focus
($<10$\% degradation of the PSF delivered by the atmosphere and
telescope). Since fibers are positioned on the telescope focal surface
with a plug-plate system, plugging must also avoid damage to the plate
or ferrule.  We achieve all these requirements through design features
in our plug plates and ferrules, setting tolerances that balance
placement precision with ease of plugging, and a detailed accounting
of the error stack-up ensured in practice by quality-assurance
measurements of as-built components.

A plug plate consists of a 0.787~m diameter disk of 3.2~mm thick
6061-T6 aluminum drilled with precision holes to a positional accuracy
of 9~$\mu$m RMS. The IFU, sky, guider, and calibration fibers are all
plugged into these holes.  Upon installation in the cartridge the
plates are bent to match the shape of the focal surface of the
telescope, which forms a dish 2.6~mm lower in the center than at the
field edge \citep{Gunn+2006}. This focal surface is non-telecentric.
To ensure the fibers are aligned with the incoming ray bundle, the
plug plate holes are drilled while bent over a mandrel with suitable
curvature.  The maximum angle which the vertex of a ray-bundle makes
to the normal of the plate is found at $\sim$2/3 of the field radius,
corresponding to 0.035 radians, or $\sim$2 degrees. The maximum vertex
deviation is 35\% of the half-angle input cone at f/5, and
consequently the proper angling of the bore holes is important to
avoid so-called non-telecentric FRD \citep{Wynne1993}.

\begin{table}
\caption{MaNGA Plugging Positional Error Budget\label{instrument:tab:plugging_error}}
    \begin{tabular}{lccl}
      \hline Error sources & \multicolumn{2}{c}{Error (mm)} & Measurement \\ \cline{2-3}
      & As-built & Measured & method\\
      \hline \hline
      Ferrule concentricity & 0.009 & 0.003 & Smartscope\\
      Plugging OD tolerance &0.004 &0.004 & micrometer\\
      IFU assembly clearance &0.015 &0.003 & SDSS test-stand\\
      Plugging hole position &0.009 &0.005 & CMM\\
      Radial hole clearance &0.014 &0.014 & none\\
      {\bf Total RMS error} & {\bf 0.024} & {\bf 0.016} & \\
      \hline
    \end{tabular}
\end{table}

\begin{figure}[tb]
  \centering
  \includegraphics[width=0.45\textwidth]{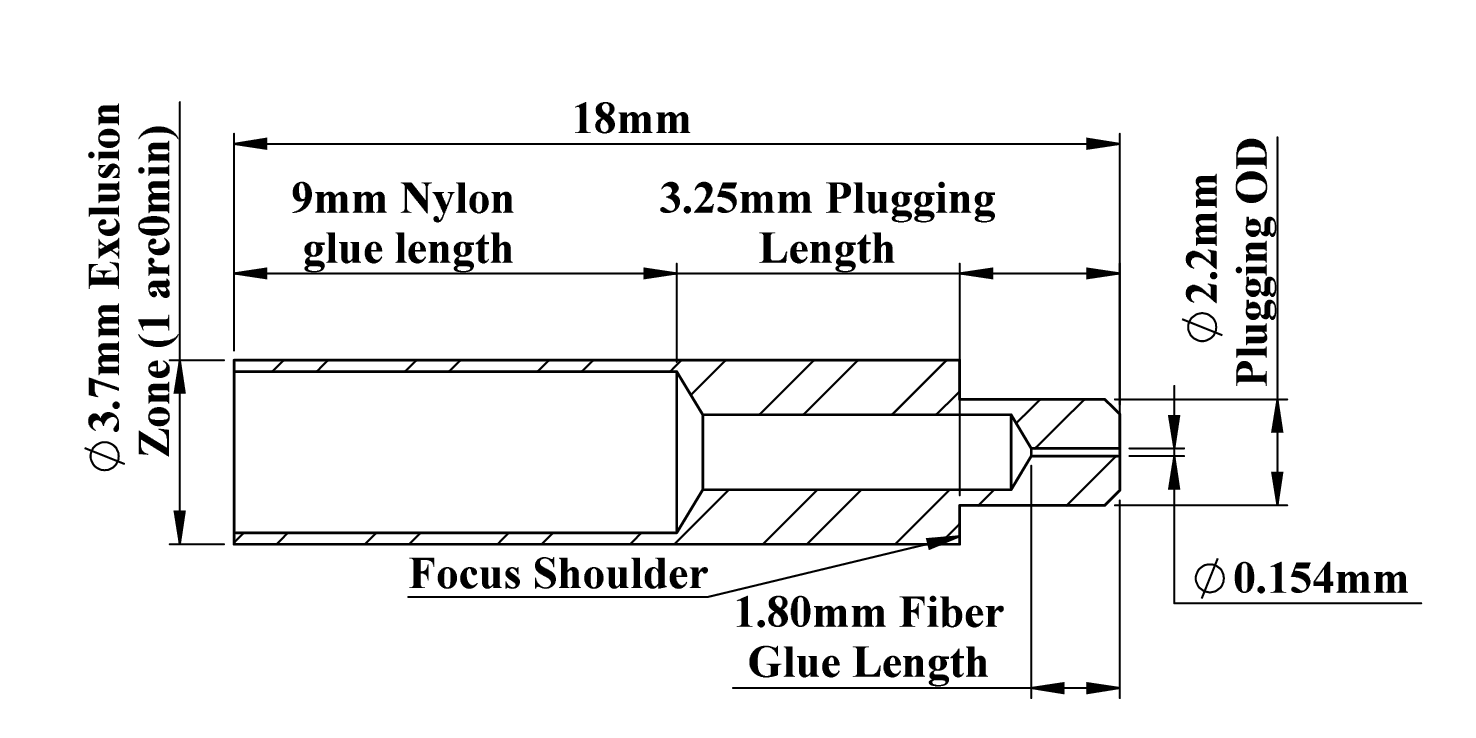}
  \caption{Cross section and isometric view of a sky ferrule
    with key dimensions shown. Each sky ferrule contains a single fiber
    on the central axis.
  \label{fig:sky-ferrule}}
\end{figure}

The on-sky orientation of the IFUs is set by a clocking pin in the IFU
ferrules, matching a separate smaller hole drilled into the
plug plate.

The focal distance of each ferrule is set by a shoulder on the ferrule
which acts as a stop against the back side of the plate (see
Fig.~\ref{fig:ferrule-cross} and \ref{fig:sky-ferrule}). In optimizing
the focal distances we have considered the consequences of
compensating for the non-telecentric beam and curved focal surface on
the extended, but flat IFU surface. There are two issues here.  The
first concerns the simple fact that the IFUs have a flat surface.
Clearly, the larger the IFU fiber surface the greater the deviation
from the curved focal surface at the edge of the IFU. This is
compounded by the requirement to telecentrically align the IFUs. In
the worst case the IFUs must be tilted by 2 degrees with respect to
the focal-surface normal for telecentric alignment. In this case the
penalty in image quality degradation due to defocus in an f/5 is
$\sim$0.25$\asec$ for the larger IFUs (the edges of the IFU are
$\sim$31~$\mu$m above/below focus if the center is in focus), which
meets our requirements.

\begin{figure}[tb]
  \centering
  \includegraphics[width=8cm]{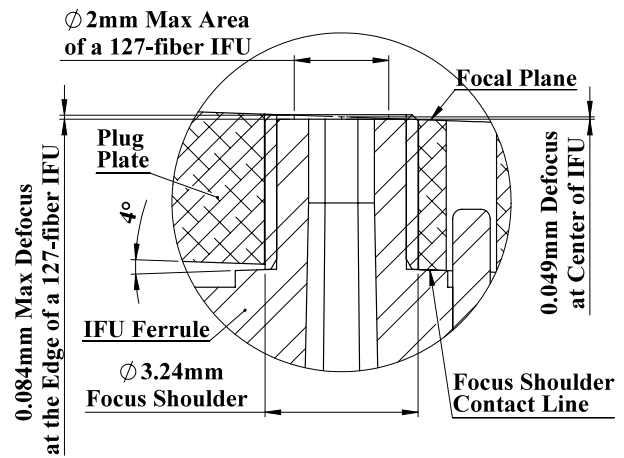}
  \caption{Schematic of ferrule contact with the plug plate showing
    (a) the defocus caused by the curved focal plane and requirements
    for telecentric alignment of the IFUs; and (b) the tapered focus
    contact shoulder that mitigates defocus due to the curved plate
    and angled plugging holes.
 \label{fig:hex-ferrule-defocus}}
\end{figure}

The second focus issue concerns the proper pistoning of the ferrules
when they are inserted into the plug-plate. The combination of plate
curvature and hole angle (to provide telecentric alignment) modulates
the contact-point between the plate back and the ferrule shoulder, and
changes the plugging depth as a function of field position.  A simple
and effective way to mitigate this defocus is to put a slight taper on
the shoulder forming the flange at the foot of the ferrule. By doing
this, the diameter which determines the vertical intrusion (focus) is
the diameter of the ferrule tip rather than the outer diameter of a
contact surface, which can be smaller by a significant amount (see
Fig.~\ref{fig:hex-ferrule-defocus}).  Despite being a single point of
contact at the lower side of the bore, the focus error can be reduced
to the point where it is no longer significant relative to the
seeing. At the edge of our largest IFU this translates to a
0.28$\asec$ image degradation although the typical defocus is
$\sim$0.16$\asec$.

To determine acceptable ferrule clearance, a set of plates was built
with different plugging interface tolerances.  These were tested to
balance the ability to achieve reliable plugging without excessive
force on the ferrule while maintaining positioning and clocking
accuracy in accordance with our requirements.  An optimal plugging
hole clearance of 20~$\mu$m (diameter) with a larger clocking pin
diametric clearance of 172~$\mu$m was found to be the tightest
practical tolerance acceptable for plugging. These clearance
tolerances lead to positional errors of 22~$\mu$m (0.37$\asec$) with
$\pm$3~degrees of rotational error.

The majority of the errors generated in building plates and IFUs are
measurable \textit{post facto}. For example, we can measure the
as-built fiber positions in the IFUs as well as the plugging holes in
the plate. Table~\ref{instrument:tab:plugging_error} lists these
errors.  The first three lines give errors associated with the ferrule
fabricated dimensions and the final placement of the fibers within the
ferrule aperture.  The ``Ferrule concentricity'' measures the
centering of the hexagonal ID to circular (plugging) OD. The ``IFU
assembly clearance'' is the offset of the as-assembled hexagonal
array of finers with respect to the ferrule hexagonal ID (errors on
individual fiber locations, given in \ref{sec:packing-regularity}, are
part of this component).  The fourth line gives the error associated
with the location of the holes drilled in the plug-plates. ``As-built
error'' refers to what we expect for fabricating
tolerances. ``Measured error'' refers to how well we know the as-built
dimensions after using the measurement-method given in the last
column. The total RMS error refers to the uncertainty of where fibers
will be placed on sky. As-built errors stack up to 22~$\mu$m, or
0.37$\asec$, but with our measurement process we will be able to
reduce this to 12~$\mu$m, or 0.2$\asec$.

The above error budget is for a single plugging. Note that our
measured error budget is dominated by the clearance between the plate
and the plugging ferrule. This means that if we re-plug plates between
two successive dark runs to complete a plate we increase the
positional error budget by 30\%.  In other words, there will be
0.17$\asec$ uncertainty (10\% of a fiber diameter) in the IFU
location, bundle to bundle, between two different pluggings of the
same plate with the same IFUs. In addition to the positional error
induced from repluggings a rotational error of $\pm$3 degrees due to
slop in the IFU pin hole can also be accumulated.  However, replugging
of the same plate is rare ($\sim$5\%), since it is only necessary if a
plate cannot be completed during a single dark run.

\begin{figure*}[tb]
  \centering
  \includegraphics[width=0.75\textwidth]{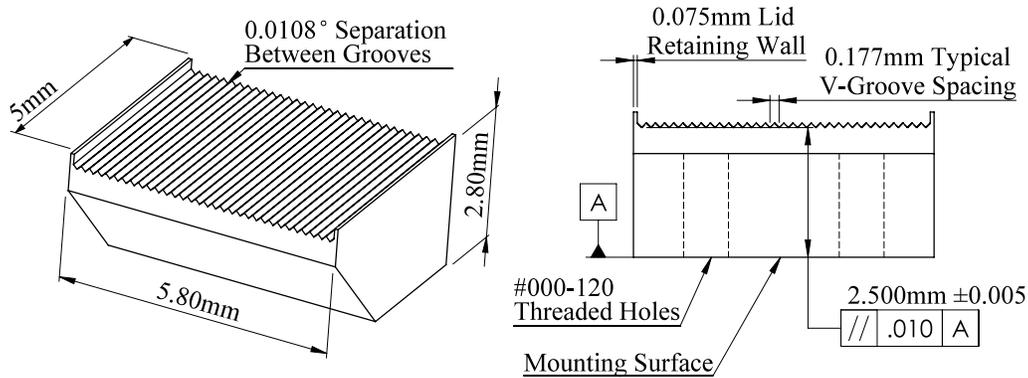}
  \caption{V-groove block for 32 fibers shown with key features
    marked. Left: isometric view of bare V-groove block with external
    dimensions.  Right: face-on view of V-groove block showing groove
    spacing, wall thickness, and high tolerance relationship between
    mounting face and the V pattern.\label{fig:vgroove-block}}
\end{figure*}

\subsubsection{Sky Ferrule Design}
\label{sec:sky-ferrule-design}

Sky-fiber ferrules (Fig.~\ref{fig:sky-ferrule}) are nearly identical
to BOSS and APOGEE ferrules (\citealp{Smee+2013}, Wilson et al., in
prep). The ID is reduced to accommodate MaNGA's smaller 151~$\mu$m OD
fiber. By using the same overall ferrule design for the sky fibers we
are able to use similar tooling and machine code for production and
verification measurement, and keep the majority of holes on a MaNGA
plate the same size as the APOGEE plugging holes, further reducing
production costs. Sky ferrules plug into precision-drilled holes on
the plug-plate with 25~$\mu$m of diameter clearance. The shoulder of
the ferrule sets field focus by registering on the bottom of the
plug-plate.

\subsection{V-Groove Blocks}
\label{sec:v-groove-block}

Blocks come in five different groove (fiber) counts and two different
groove spacings. Blocks are built to contain 21, 32, 33, or 39 fibers
with 177~$\mu$m spacing, from which any of the 5 harnesses can be
built; and to contain 24 fibers with 204~$\mu$m spacing for the
mini-bundles.  Each V-groove block consists of a 17-4 stainless steel
base and lid, with a series of 90~deg (i.e., V-shaped) grooves
machined into one side with a 100~$\mu$m EDM wire. These are tiny,
high-precision parts (5~mm in length along grooves, a total thickness
$\sim$3~mm, and a width between 3.85~mm and 7.05~mm depending on fiber
count).  Several critical design elements are incorporated relative to
the initial BOSS design based on lessons learned in APOGEE (Wilson et
al., in prep), which was built after the original BOSS fiber system
was completed.  Modifications include the design of the V-groove mount
to the slit-plate, and corrections to the fan-angle of the grooves to
account for refraction at the fiber-air interface.
Fig.~\ref{fig:vgroove-block} gives an overview of the V-groove block
design.

Block thickness is larger than for BOSS to allow \#000-120 taped holes
for attachment to, and possible later removal from, the slit
plate. The additional thickness of 1.25 mm or a total of 2.4 mm
amounts to negligible light-loss from obstruction in the collimated
beam of 1.5\%, and enables much easier assembly and replacement. In
comparison, the original BOSS slit blocks are affixed via epoxy.

The grooves fan out slightly so that the output beam of the fiber is
matched to the spherical focal surface of the BOSS spectrograph's
collimator. The radius of curvature of the slit is 640.1~mm
\footnote{This value corresponds to $R/2$ of the as-built collimator,
  and deviates slightly from the value quoted in \citet{Smee+2013}},
but the fibers in a block are polished flat. Taking into account the
n=1.46 index of refraction of the fused-silica fiber requires a
V-groove fan-angle of 0.0108~degrees for an end-spacing of 177~$\mu$m
(or a radius of curvature of 935.9~mm) for proper illumination of the
collimator. Because of the flat block face the fibers do not all sit
exactly on the focal surface.  However, for the largest block this
amounts to {$\pm$}5~$\mu$m of defocus, or only $\sim$1.2~$\mu$m of
image blur in an f/4 beam, which is negligible compared to the
120~$\mu$m fiber core diameter.

The alignment specifications on the V-groove blocks for tip, tilt and
piston are satisfied through a combination of manufacturing tolerances
on the V-groove blocks, the slit plate (see \ref{sec:slit-plate}), and
the installation procedures. The specifications and manufacturing
tolerances for the V-groove blocks are given here; the total stack-up
or slit assembly is discussed in \ref{sec:v-groove-block-alignment}.

The blocks are precision ground such that their flatness and
perpendicularity ensures seating on the slit-plate that reduces
out-of-plane (tip, or dispersion-axis) angular misalignment limited by
the uniformity of the groove depth. The out-of-plane groove tilt is
required to be {\textless}10~$\mu$m across the 5~mm block length, or
{\textless} 0.11~degree. This is the same specification as adopted by
BOSS and APOGEE. For the BOSS collimator this requirement keeps the
beam deviation at the collimator under 1\%. The tip and tilt error for
the plane of polish is 0.5~degree. This is controlled by tooling
during polish at C-Tech. The alignment of the blocks in tilt and
piston on the slit-head are determined during installation, as
discussed below.

The end walls of the V-groove block, which retain the lid, and the
groove peaks themselves are highly susceptible to damage from
handling.  In addition, any contamination in the V-grooves will lead
to stress on the fibers when the lid is affixed, possibly leading to
FRD and fiber mis-alignment in the groove. To minimize damage and
contamination, the vendor of the V-groove blocks, Oxford University,
developed a "no touch" manufacturing method. Each V-groove blank is
installed onto the EDM machine using a System 3R macro coin-and-part
chucking system from Renishaw Inc. This system allows for the the
V-groove block to be transitioned from the EDM that cuts the profile
of the block, including the indexing surface, to the mill for addition
of the tapped holes, and then back to the EDM for cutting the V-groves
with no direct handling of the components by the machinist.  Published
chucking repeatability of this system induces a maximum error of
5~$\mu$m. We find that the blocks are built to better than 3~$\mu$m
accuracy. A handling bar, attached after the threaded holes are added,
is built to interface with visual measuring systems at Oxford and
U. Washington so that full quality assurance can be conducted on the
blocks without any direct handling. This manufacturing method has
eliminated scrap due to handling and enabled us to mass produce high
accuracy components at low cost.

\begin{figure}[tb]
  \centering
  \includegraphics[width=8cm]{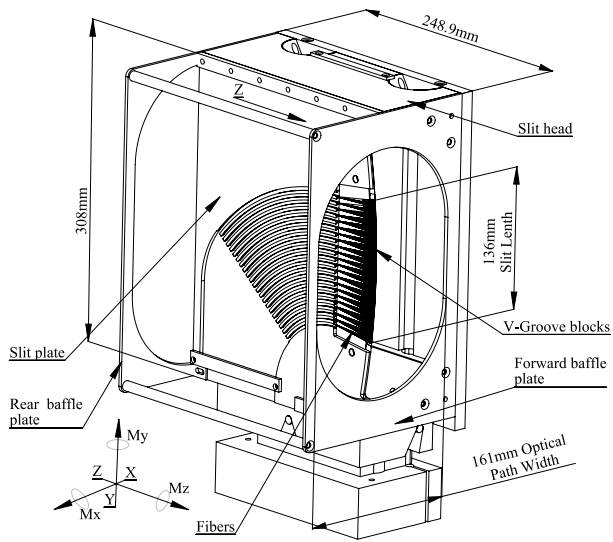}
  \caption{Schematic of slit plate installed on the cartridge slit
    head. The optical beam emanates from the fibers in the v-groove
    block in the positive $Z$ direction, encounters the collimator
    mirror surface located 640.1 mm from the fiber faces, and is
    redirected back through the baffle plates towards the camera
    optics. The slit plate obscures 3.4\% of the
    beam.\label{fig:slitplate}}
\end{figure}

\subsection{Slit Plate Assembly}
\label{sec:slit-plate-assembly}

Every V-groove block in a cartridge is bolted onto one of two slit
plates (12 total for all 6 cartridges), machined at U. Wisconsin.
Each slit plate makes up the pseudo-slit of one of the two BOSS
spectrographs. The plates act to align the blocks with the input focal
surface of the spectrographs, and retain the first 300~mm of fiber so
that the fiber and harness tubing do not interfere with the optical
path. The primary challenges in the slit plate design and assembly
include (1) minimizing the total thickness so that optical obscuration
is minimized (the slit sits in collimated beam); (2) ensuring the
plate and assembly process yield uniform fiber alignment so that the
line-spread function (LSF) is also uniform (this is particularly
important for sky-subtraction; see \citealp{Law+2014,Yan+2014}; yet (3)
providing a means for attaching and unmounting harnesses should they
need servicing.

\subsubsection{Slit Plate Fabrication Tolerances}
\label{sec:slit-plate}

Slit plates are each machined from a single 6.3~mm thick cast aluminum
6061-T6 plate measuring $\sim$300~mm high by $\sim$240~mm deep shown
in Fig.~\ref{fig:slitplate}. Each plate contains T-shaped channels for
each V-block set of fibers that help guide and retain the fibers while
allowing for thermal expansion and ease of assembly. Slit plates also
feature precision-machined mounting surfaces for the V-groove blocks.
Front surfaces of the plates are angled at 45 degrees to minimize
internal reflections entering the spectrographs.

To limit the amount of beam obscuration, the cross section of the
plate in the area of V-groove block mounting is only 1~mm, the same as
for the existing BOSS plates. (As discussed in
\ref{sec:v-groove-block}, the vignetting is dominated by the V-groove
blocks.)  A key manufacturing concern is the variability of this
surface due to its thinness, compounded by the large amount of
material removed (84\%) from the original cast aluminum
plate. Deviation in the height ($X$) or tilt ($M_y$) of the surface,
corresponding respectively to slit-position and input angle shifts in
the spectral dimension, can lead to variations in the spectrograph
LSF. These LSF variations would be superimposed on an otherwise
smoothly changing LSF due to field-dependent aberrations in the
spectrograph optics.

Alignment variations between blocks are of more concern than the total
variation across the slit length. Our ability to {\it accurately}
subtract sky with the highest precision requires interpolating between
as many sky spectra as possible, sorted by some measure of their LSF
and locality on the sky, as discussed in
\citep{Law+2014}. Discontinuous changes in input angle or height from
fiber to fiber or block to block translate into discontinuous changes
in the LSF for sky fibers that are close together on the sky. To
ensure the slit plates will not induce any large LSF variations, each
plate is measured using a coordinate measuring machine (CMM) at U.\
Washington. The global threshold for displacement (i.e., deviations from a
perfectly flat surface) in the spectral direction ($X$) is set at 100
$\mu$m (displacements of 40 $\mu$m along the slit are typical) with an
allowable tilt in the spectral direction ($M_y$) of 0.1 degree between
block mounting surfaces. Linear trends in ($M_y$) are removed by
shimming the slit plate at the time of installation; higher order
trends are negligible, resulting in an rms of 0.02 degree.

\begin{figure}[tb]
  \centering
  \includegraphics[width=0.45\textwidth]{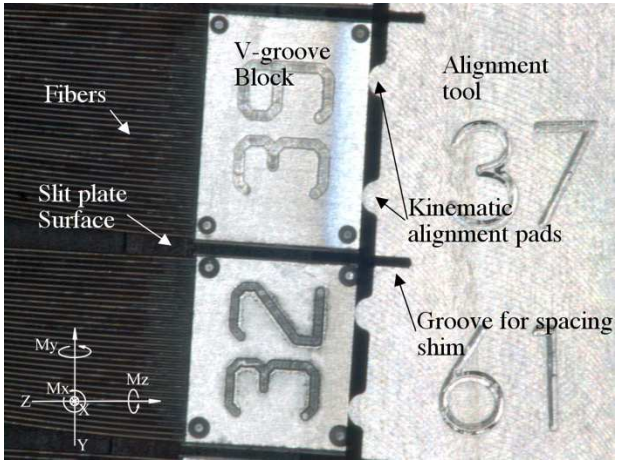}
  \caption{Image of a portion of the slit with the alignment fixture
    in place. The features that control the 6~degrees of freedom
    (translation and rotation) are indicated.  The numbers cut into
    the tops of the V-Groove blocks denote the number of fibers
    contained within the blocks.  These are primarily used for
    component identification during assembly.  The text on the
    alignment tool guides the order of block assembly.
    \label{fig:axis}}
\end{figure}

\subsubsection{V-Groove Block Alignment}
\label{sec:v-groove-block-alignment}

The final phase in the slit-plate assembly is the attachment and
alignment of the V-grooves onto the slit plates. Two small $\#$000-120
threaded holes on the bottom are used to attach the V-groove block to
the slit plate. This allows a single IFU to be replaced and for its
position to be adjusted without affecting other IFUs. A 303 stainless
steel alignment tool is attached to the front of the slit plate by two
locating holes. The alignment tool serves as the reference surface to
which V-groove blocks are positioned in focus (Z), tip in the spatial
direction ($M_x$), and position (Y) along the slit. The V-groove block
attachment screws are tightened to establish position and eliminate
rotation in the spectral direction (X and $M_y$, respectively), and to
remove any yaw ($M_z$) in the blocks with respect to the spatial axis
of the slit. Controlling these latter three degrees of freedom relies
on the flatness of the V-groove block and slit plate surfaces, as
discussed in \ref{sec:v-groove-block} and \ref{sec:slit-plate},
respectively.

Focus and tip are fixed by pushing the face of the V-groove block up
against two kinematic indexing features on the face of the alignment
tool.  The tangent points of the kinematic features are on a surface
with a 640.1~mm radius of curvature to position the fibers as closely
as possible on the collimator focal surface.  Since the V-groove block
face is flat, at most two of the fibers can be perfectly focussed on
this surface.  To minimize defocus along the V-groove, every V-groove
block contacts two of these features 1/3 and 2/3 along the face length
of the block (Fig.~\ref{fig:axis}). This configuration keeps the
maximum defocus to 5~$\mu$m which is negligible given the
approximately f/4 exit beam of the fibers and their 120~$\mu$m core
diameter. The contact of the V-groove block with the kinematic
features on the alignment tool is slightly below the fiber faces (to
avoid fiber damage) but still on the same polished surface as the
fibers. The position along the slit is controlled with 280~$\mu$m
shims built into the alignment tool that interlock with the gaps
between the V-grooves to position them to within 25~$\mu$m of their
intended position. The nominal distance between neighboring V-groove
blocks is 313~$\mu$m.

Table~\ref{slit-build-errors} lists the accuracy of the as-built slit
assembly. The accuracy achieved is close to that predicted by the
tolerance analysis.  The shift in X is slightly worse than anticipated
although it has no effect on the performance of the instrument.  The
defocus of up to $\sim$8~$\mu$m also has little effect corresponding
to an increased images size of $\sim$2~$\mu$m.

\begin{table}
\caption{\label{slit-build-errors} Slit Build Errors }
  \begin{center}
    \begin{tabular}{ccccc}
      \hline
&Spectral&Focus&Tip in Fiber&Tip in Spectral\\
&Direction(X)&(Z)&Direction(M${_x}$)& Direction(M${_y}$)\\
      & (mm) & (mm) & (degrees) & (degrees) \\
     \hline
      \hline
      Predicted & 0.045 & 0.018 & 0.28 & 0.035\\
      Measured & 0.060 & 0.008 & 0.05 & 0.05\\
     \hline
    \end{tabular}
  \end{center}
\end{table}

\subsubsection{V-Groove Ordering in the Pseudo-Slits}
\label{sec:v-groove-ordering}

Because the total number of IFUs is odd, and because of the range of
block lengths required for the different IFUs, some asymmetries in the
layout of the slits is requred. Tables~\ref{spectrograph1-ifu} and
\ref{spectrograph2-ifu} give the order and description of IFUs along
each of the two slits, and the amount of slit length used to
accommodate each IFU. The location of the fibers for each IFU along
the slit is given for the central fiber (in the slit) with respect to
the optical axis. At 69~mm from the optical axis the collimator stop
vignettes the beam by $\sim$5\%. This location was used as a practical
limit for the available slit length.

Considering the 138~mm of total slit length per spectrograph, our IFU
complement, and fiber spacing, we attempted to optimize the slit
layout with the following desiderata. First, all V-groove blocks
associate with a single IFU were placed contiguously along the slit.
Another priority was given to positioning the mini-bundles' V-groove
blocks as close to $1/3$ and $2/3$ along the slit length to best
sample the slit with calibration standards.  The design of the mini
bundle is such that it can be swapped into the position of any 61
fiber IFU. This was done so that, if needed in the future, four more
calibration mini-bundles could be added to each existing slit at the
cost of two of the four 61-fiber IFUs.  In order to allow for this
swap, the two replaceable 61 fiber IFUs had to be placed at the ends
of the slit, so that the individual mini-bundle V-groove blocks, which
are significantly larger than the 32 fiber V-groove blocks used on the
61 fiber IFU, would fit on both sides when installed in place of the
32 fiber block from the IFU harness.

\begin{table}
\caption{\label{spectrograph1-ifu} Spectrograph 1 slit order}
\begin{center}
\begin{tabular}{cccc} \hline
 IFU Type & Number of & Slit Position of & Slit Length\\
 & V-Groove Blocks & Central Fiber & Used\\
      & & (mm) & (mm)\\ \hline \hline
      61 fiber IFU & 2 & 62.09 & 12.1\\
      127 fiber IFU & 4 & 43.04 & 25.4\\
      Mini Bundle & 1 & 27.54 & 5.0\\
      37 fiber IFU & 1 & 21.21 & 6.9\\
      61 fiber IFU & 2 & 11.34 & 12.1\\
      127 fiber IFU & 4 & -7.7 & 25.4\\
      91 fiber IFU & 3 & -29.8 & 18.5\\
      Mini Bundle & 1 & -41.71 & 5.0\\
      37 fiber IFU & 1 & -48.05 & 6.9\\
      61 fiber IFU & 2 & -57.92 & 12.1\\
      19 fiber IFU & 1& -66.22 & 3.7\\ \hline
\end{tabular}
\end{center}
\end{table}

\begin{table}
\caption{\label{spectrograph2-ifu} Spectrograph 2 slit order}
\begin{center}
\begin{tabular}{cccc} \hline
IFU Type&Number of & Slit Position of & Slit Length\\
 & V-Groove Blocks & Central Fiber & Used\\
     & & (mm) & (mm)\\ \hline \hline
      61 fiber IFU & 2 & 62.09 &12.1\\
      127 fiber IFU & 4 & 43.04 & 25.4\\
      Mini Bundle & 1 & 27.54 & 5.0\\
      37 fiber IFU & 1 & 21.21 & 6.9\\
      91 fiber IFU & 3 & 8.28 & 18.5\\
      127 fiber IFU & 4 & -13.92 & 25.4\\
      91 fiber IFU & 3 & -28.74 & 18.5\\
      Mini Bundle & 1 & -48.98 & 5.0\\
      37 fiber IFU & 1 & -48.05 & 6.9\\
      127 fiber IFU & 4 & -51.49 & 25.4\\
      37 fiber IFU & 1 & -57.92 & 6.9\\  \hline
\end{tabular}
\end{center}
\end{table}

\subsubsection{Stray Light Considerations}
\label{stray-light}

The MaNGA slit obscures $5.2\times 160$~mm of a 160~mm diameter beam,
or 3.4\% of the light, assuming a uniform beam profile. A rectangular
baffle at the center of each collimating mirror masks the light which
would be reflected onto the slit plate and V-groove block assembly, to
minimize ghosting, scattering, and stray light. Because the baffle was
designed for the somewhat thinner BOSS slit-plate assembly, there was
some concern that ghosting from reflections might be present despite
the majority of front-facing surfaces being angled at 45
degrees. Hence, the slit plate is anodized with a standard, unsealed,
type-II black anodizing and subsequently coated with Aeroglaze Z306.

To quantify the effect of stray light and possible ghosting, we tilted
the collimator through extreme angles in the spectral direction and
obtained over-exposed arc-lamp frames.\footnote{Because each cartridge
  places the slits in a unique (but highly repeatable) location with
  respect to the BOSS spectrograph collimators, these mirrors can be
  actuated in tip, tilt, and piston to adjust the position of the
  spectra on the CCD.}  At the optimal tilt position of the collimator
there was no detectable reflected light from the side of the slit
plate or reflections from the front of the V-groove blocks, suggesting
that the currently installed collimator central baffle works well for
MaNGA's slightly thicker slit assembly.

\subsection{AR Coating}
\label{sec:ar-coating}

The Fresnel loss from each air-glass interface with the fused-silica
(FS) fibers amounts to roughly 4\% at visible wavelengths. Significant
additional losses can come from surface-roughness, due to imperfect
polishing, which leaves blemishes on physicals scales of order a
wavelength (0.5~$\mu$m). These imperfections also cause scattering
that contributes to FRD. A common solution to both problems is to bond
the fiber ends with optical gel to an AR-coated FS plate. This has the
advantage of providing a protective cover, and in principle reduces
the required level of fiber polish quality.  However, the application
on the small scale of our IFUs is mechanically challenging.

Another approach is to AR coat the fibers themselves. Direct coating
has proven difficult in the past, however Cascade Optics has recently
developed a high-performance, multi-layer coating process applied at
temperatures and vacuum pressure levels that will not damage the IFUs
(epoxies, polyimide fiber buffer, or fiber jacketing). Contamination
from outgasing of harness material in the deposition chambers remains
a challenge for guaranteeing the best possible performance, but
delivered performance, as verified by testing witness samples and the
harnesses themselves, demonstrates significant throughput gains.

Unlike most AR coating applications, the AR coating on the fibers
covers the fiber faces, the epoxy of the IFU and v-groove termination,
and the stainless steel surroundings, all of which have different
thermal properties.  Because of this CTE mismatch, less than optimal
coating parameters were used, namely yet lower vacuum and substrate
temperature. Our specification for the coatings are (1) to achieve a
reflectivity of {\textless}1.5\% in the wavelength range of
350-1000nm; (2) during coating, the maximum application temperature
must not exceed 50~C, with a minimum pressure of 1e-3~Torr; and (3)
during use, coatings must withstand temperature variations from -6~C
to 21~C (500 cycles) and humidity variations from 0-100\%~RH (500
cycles) with no measurable change in optical properties. To test
durability (adhesion), coated fiber cables were cycled 500 times and
showed no signs of mechanical or optical degradation.

\begin{figure}[tb]
  \centering
  \includegraphics[width=8cm]{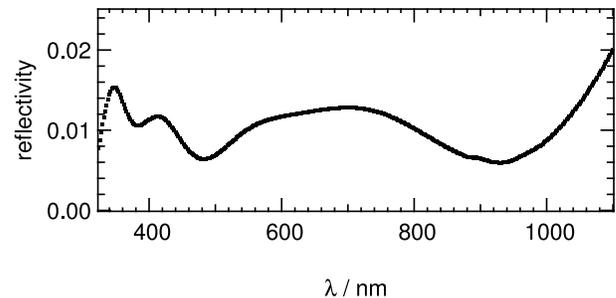}
  \caption{Reflectivity measurement of the AR coating as applied to a
    witness sample.\label{fig:AR-reflectivity}}
\end{figure}

The reflectivity curve for a witness sample is shown in
Fig.~\ref{fig:AR-reflectivity}. The variations from batch to batch are
at the 0.5\% level between 350 and 950 nm.  We find the mean
throughput increases by 4.5\% (absolute) from 91$\pm$1.1\% to
95.8$\pm$0.3\%. Both the increase in the mean throughput and the
decrease in throughput variation is significant. We attribute the
latter to the improvement in the surface smoothness (e.g.,
\citealp{Eigenbrot+2012}).

Our measurements were made primarily at 550~nm (in a band-pass defined
by the Johnson~V filter). To verify the broad band-pass of the coating
several fibers were measured at 420~nm and 800~nm in intermediate
(Stromgren) and broad (Johnson~I) filters, leading us to conclude that
the throughput values from 420-800~nm are uniform (see
Fig.~\ref{fig:FBP-transmission}). In summary, these AR coatings
improve our harness throughput from 89-92\% to 95-96\% (4.5\% gain in
the mean) with manageable risk given our experience with life-time
testing and thermal cycling.

\subsection{Strain Relief and Stress Considerations}
\label{sec:strain-relief}

Fiber stress can lead to FRD-induced throughput loss, and arises in our
system from three principle sources. The first is a combination of
epoxy and polishing-induced damage and surface roughness at the fiber
terminations. For this reason, the fiber is only attached to the
assembly at the small glue joint at the ferrule tip and again at the
V-groove block attached to the slit of the spectrograph. These
contributions to fiber stress appear to be minimal, as indicated by
our laboratory throughput measurements, due to the minimal glue length,
careful handling, and AR-coating.

The second source of stress arises from the routing of the harnesses
in the cartridge and their handling during plugging and storage.
Fiber jacketing is attached to both sky and IFU ferrules with a glue
joint on the back end of the ferrule. This acts as a strain relief to
insure that load induced during plugging operations is not transmitted
to the fiber. The fiber jacketing over the 0.7~m of the plugging
length of the IFU is rugged PVC covered steel mono-coil tube.  Inside
of this tube all of the fiber are bundled into a soft Teflon tube,
which slides easily inside of the mono-coil.

The third source of stress can arise due to CTE mismatch between the
FS fiber and the harness jacketing. Differential changes in length
over the 1.8~m harness between -6~C and 21~C are estimated to be no
more than 2.4~mm.  Roughly 45~mm of bare fiber extend between the
end of the V-groove block jacketing and the back of the V-groove
block.  The jacketing and fiber are carefully routed in T-shaped
slots in the slit-plate (\ref{sec:slit-plate}). This length is more
than ample to decouple the low CTE fiber and high CTE jacketing.

\section{Metrology}
\label{metrology}

Metrology throughout the production process played an important role
in the development of the IFUs and in their subsequent observational
use. This included design-contingent measurements, such as the
as-built OD of the fiber for fine tuning of the ferrule ID;
measurements for quality-assurance and acceptance of components on a
pass/fail basis, such as sky ferrule concentricity and V-groove blocks
flatness; and ending with as-built measurements that are used in data
reduction to better constrain the positional accuracy of the IFUs on
sky. Metrology data used for data reduction must be readily
available to the data reduction pipeline.  To store and access all of
this information, a database has been developed called {\tt mangaCore}
\citep{Law+2014}.  This database includes all of the information
derived during production.

\subsection{Hex Ferrules}
\label{sec:hex-ferrules}

After machining the ferrules, we use a SmartScope video measuring
system built by Optical Gauge Products, Inc., to measure all of the
critical dimensions of the hexagon 10 times and ensure that they are
within $\pm$3~$\mu$m of the intended size.  Non-complient ferrules are
discarded. The rotation of the hexagon relative to the pin is also
measured and recorded, as is the relative offset of the hexagon to the
ferrule OD.  Each ferrule is given a 4-digit serial number. All
measurements are ingested into {\tt mangaCore} and used in data
reduction.

\subsection{Sky Ferrules}
\label{sec:sky-ferrules}

Sky ferrules are measured in a similar way to the hexagonal
ferrules. The most critical dimensions and uncertainties are as
follows: ferrule ID, 0.154~$+$0.004/$-$0.003~mm; concentricity with
respect to the OD, 0.013~mm; and ferrule plugging OD,
2.154$\pm$0.003~mm. The resulting positional uncertainty in the sky
fiber location from all error sources is accounted for when selecting
targeted sky positions.

\subsection{V-Groove Blocks}
\label{sec:v-groove-blocks}

Several important dimensions of every V-groove block are also
repeatedly measured to ensure that the blocks adequately position
fibers with respect to the spectrograph.  The groove spacing must be
0.177$\pm$0.005~mm for the IFU block and 0.20$\pm$0.005~mm for the
mini-bundles.  Measurements of the V-groove positions at the back in
addition to the front surface of the blocks can is used to ensure that
the fan angle of the block is 0.0108~degrees. The height of the
V-groove block above the indexing face is controlled to
2.500$\pm$0.005~mm with a planarity requirement of 0.010~mm. As with
the ferrules, automated analysis discards non-complient parts.

\subsection{Plug Plates}
\label{sec:plug-plate}

The plug plate verification system is largely unchanged from earlier
incarnations of SDSS.  The positions of the IFU holes on the plates is
measured on the Coordinate Measuring Machine (CMM) at U.\ Washington.
Their positions are required to be within 12~$\mu$m RMS of the
designed location. The hole locations are then incorporated into {\tt
  mangaCore} for use in data reduction.

\begin{figure*}[tbh]
  \centering
  \includegraphics[width=0.75\textwidth]{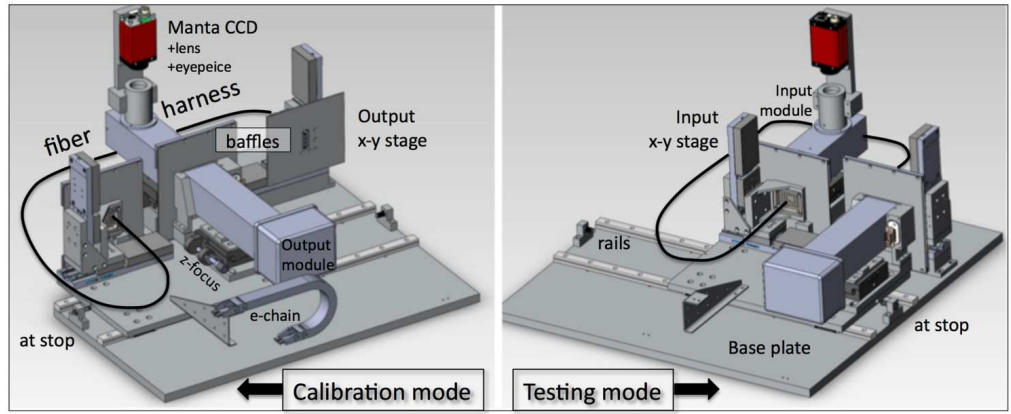}
  \caption{Schematic diagrams of SDSS test-stand in testing mode
    (right) and calibration mode (left). Light purple units are
    heritage input and output modules. The red unit is the CCD camera
    used for machine-vision control of the input stages and IFU
    metrology.  Not shown, for clarity, are a post for manual movement
    of the output module sub-stage, the source lamp (situated next to
    the input module in the cut-out base plate corner), fiber source
    connection from the lamp to the rear of the input module, and all
    of the electonic cabling connecting to a separate electronics
    module. \label{fig:tester-layout}}
\end{figure*}

\subsection{Laboratory Measurements of IFUs}
\label{sec:laboratory-measurements-of-ifus}

Three fiber test-stands are available to MaNGA for development and
production. The first, referred to as TSW, is lab infrastructure at
U.\ Wisconsin developed for fibers on the WIYN and SALT telescopes
\citep{Bershady+2004,Crause+2008}, capable of measuring throughput and
FRD (far-field profile) in a filtered band-pass with an adjustable
input beam. This is a double-differential system that compares direct
and fiber-fed output far-field beam images while simultaneously
monitoring the filtered input source with a photo-diode. A slide
enables rapid exchange of filters with minimal need to refocus between
400-800~nm. Recent addition of pellicles allow for near-field imaging
of both input and output fiber surfaces. By measuring the far-field
pattern it is possible to separate throughput and FRD signals. Results
using this system have been shown in previous sections for the
test-termination and prototype IFUs
(Fig.~\ref{fig:ring-throughput}). However, the system is manually
operated and therefore time-consuming for making measurements on many
fibers; it is also not designed for precision metrology measurements
of the IFU face so it is not used for production validation of the
MaNGA IFUs.

The second and third test-stands are identical units based on the
heritage SDSS test-stands \citep{Smee+2013}, one (TS1) intended for
delivery to the vendor to determine performance compliance of fiber
products before shipment, and the other (TS2) for verification at
U.\ Wisconsin after AR-coating but before installation on the
telescope.

For SDSS-IV, the test stands were augmented to handle automated
testing of all fibers in MaNGA via 4-axis, computer-controlled
actuation of the fiber input and output mounts.  In their modified
configuration, the test stands serve two primary functions: (i)
verification of throughput, with a requirement of 2\% precision and
external accuracy to avoid spurious rejection of harnesses for low
performance at about the 4~$\sigma$ level; and (ii) measurement of the
as built position of each fiber and its mapping within the IFU head,
with a precision requirement of relative fiber locations better than
2~$\mu$m. For backward compatibility, the interface and automated
control also handles single-fiber harnesses for other SDSS-IV
experiments such as eBOSS and APOGEE-2. The new test-stand layout is
shown in Fig.~\ref{fig:tester-layout}.

\begin{figure*}[tbh]
  \centering
  \includegraphics[height=1.65in]{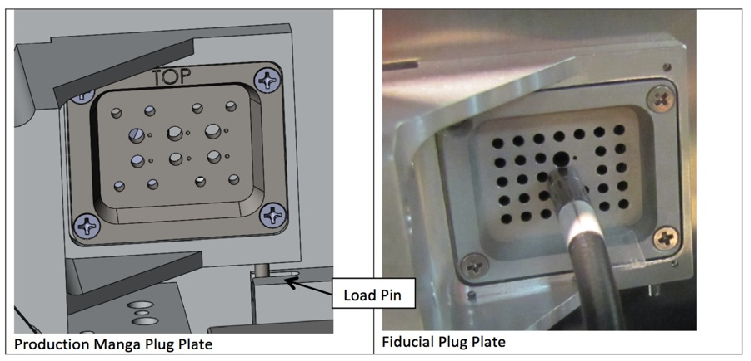}
  \includegraphics[height=1.65in]{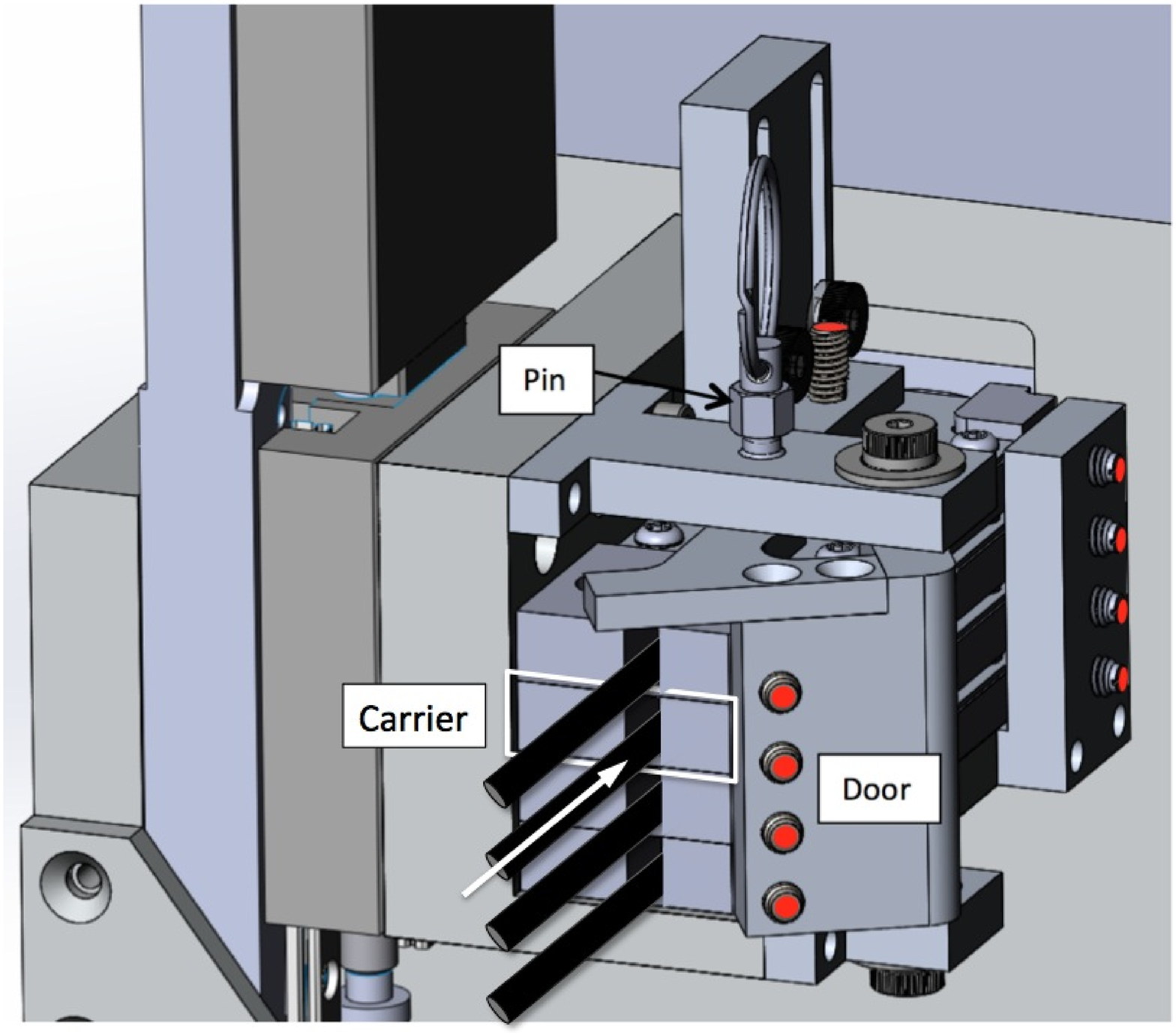}
  \caption{Test-stand fiber interface. Ferrule plug-plates for the
    input module are shown for a MaNGA calibration harness (left) and
    an IFU harness or single-fiber harness for eBOSS and APOGEE
    (center). The central panel shows a proto-type IFU ferrule
    plugged, with a second pair of holes (large and small) for the
    production IFU ferrule directly above it. A load pin (labeled)
    ensures mechanical stability of the focal position during the
    mechanical strain of plugging and unplugging.  The right hand
    panel shows the output module loading mechanism (labeled pin and
    door) for the V-groove blocks, mounted in their rectangular
    carriers described in the text. Vertical stacks of up to four
    carriers can be accommodated. Red dots indicate pre-loads, while
    the white arrow on the harness tubing emerging from the carriers
    indicates the photon path toward the output
    module.\label{fig:tester-plugplate}}
\end{figure*}

For testing, all fibers ferrules in a given harness are loaded
manually into interchangable, miniature plug-plates that mount onto
the x-y stage servicing the input module interface. The plug-plates,
shown in Fig.~\ref{fig:tester-plugplate}, are 1.5~in $\times$ 1.75~in
in size, with active plugging regions of 0.75~in $\times$ 1~in that
enable the motorized stages to center each fiber on the input beam.
One plate is configured for the MaNGA calibration harnesses,
containing three, pinned mini-IFU ferrules and eight sky ferrues. A
second plate covers all other harnesses, containing holes for a pinned
IFU ferrule and up to 30 sky-fiber ferrules (to also accomodate eBOSS
and APOGEE harnesses). At the output end of the system, the V-groove blocks
are stacked vertically on a matching x-y stage that services the
output module interface. To ensure the fiber exit surface is well
aligned with the output module (parfocal with the input beam), each
V-groove is mounted and aligned (in piston, tip and tilt) inside of a
rectangular, aluminum carrier by the vendor; this carrier also serves
to protect the termination during shipping, coating, testing, and
installation. The mechanical alignment within the carrier takes
advantage of the flat block surfaces and mounting taps. The mechanical
alignment of the carriers in the output interface is ensured with
precision stop surfaces and pre-loads in vertical and horizontal
dimensions. Ferrule (input) and block (output) mechanical interfaces
enable rapid and repeatable harness loading, and they establish
nominal relative locations for all fibers in the two focal surfaces of
the test-stand.

Like the Wisconsin test-stand, the SDSS test-stands are differential
photometers, comparing a photo-diode measurement of the input beam
signal (``calibration mode'') through the same optical path as the
fiber output beam (``testing mode''). The stability of the input beam
is monitored before and after each harness measurement.

To characterize the IFU metrology and also ensure accurate placement
of the 50~$\mu$m illumination spot onto the fiber face for throughput
testing, fiber positions within the IFU and associated sky fibers are
determined by back-illuminating the V-groove blocks of the harness
with LEDs. Images of the fiber faces are taken with a camera mounted
on the input module, adapting to the pellicle beam-splitter and
eye-peice described in \citep{Smee+2013}. The camera provides a 3.5~mm
field of view sampled at 2.86~$\mu$m, with distortions of $<$ 1.3\% at
the field edge, corrected using reference images of a calibration
grid. This is sufficient to image our largest IFUs in their entirety
and provide better than 1~$\mu$m position estimates for every fiber.

\begin{figure}[tbh]
  \centering
  \includegraphics[width=0.45\textwidth]{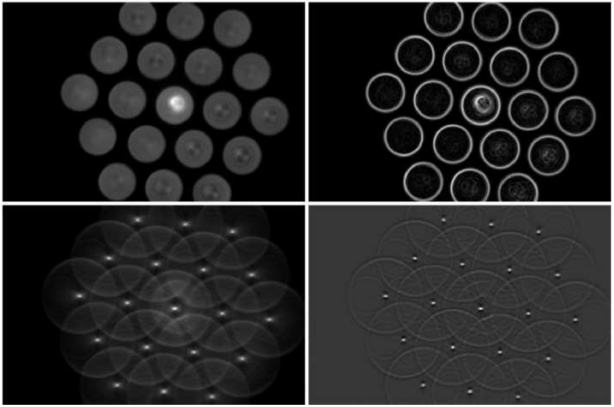}
  \caption{Representative sequence of images used to measure fiber
    metrology, as described in the text. Top-left panel is the
    original CCD image of a 19-fiber IFU, intentionally rotated for
    demonstration purposes from a fiducial orientation closely aligned
    with the CCD axes. The fibers are rear-illuminated by LEDs mounted
    to the output module. The near-field image of the f/5 input beam
    from the test-stand input module is seen within the central
    fiber. The Sobel transform (first derivative) image is at the
    top-right; the Hough transform image is at the bottom-left; and
    the final convolved image is at the bottom-right.
    \label{fig:machine-vision}}
\end{figure}

Fiber positions are measured automatically from these images via
several image transforms, illustrated in
Fig.~\ref{fig:machine-vision}. The Hough transform serves to filter
out contaminant signal in the derivative image, e.g., from the
illumination spot, while the convolution enhances the central peak
relative to the background. The fiber centers are determined from a
peak-finding algorithm on this final (convolved) image. We find the
fiber centering uncertainty to be $\sim0.3~\mu$m, well below our
requirements.

The global location of the fibers within the IFU are determined
relative to the ferrule hexagonal aperture, identified from a seprate
image of the ferrule when front-illuminated with LEDs. This centration
and clocking of this aperture with respect to the ferrule OD and pin
are known from CMM measurements of each ferrule prior to harness
fabrication, as described in Section \ref{sec:hex-ferrules}. The error
stack-up on the absolute fiber location within the ferrule is $<5~\mu$m
in translation; the rotation of the hexagonal pattern with respect to
the clocking pin is known to $<0.2^\circ$; this is well below the
22~$\mu$m translation uncertainty and $3^\circ$ rotational uncertainty
in the ferrule plugging due to the required hole tolerances for the
telescope plug plates (Section \ref{sec:plug-plage-interface}). The
metrology information is saved under the serial number of a harness in
{\tt mangaCore} for later use in correcting the positional information
in the data cube.  An example of the metrology output plots can be
seen in Fig.~\ref{fig:positional-errors}.

\begin{figure}[tbh]
  \centering
  \includegraphics[width=0.5\textwidth]{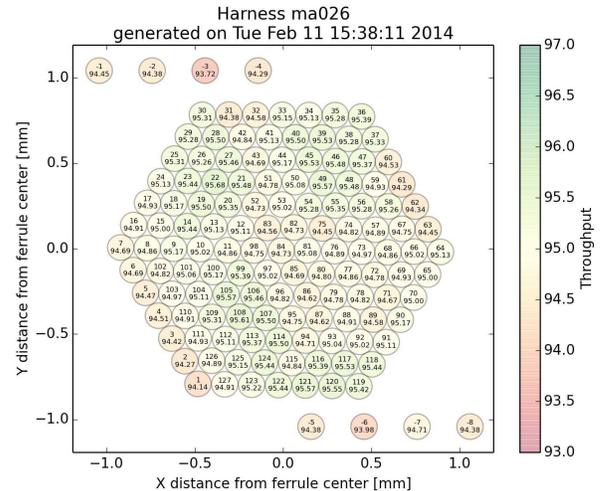}
  \caption{IFU throughput map as generated and archived for every IFU
    harness from the SDSS test-stands for a typical 127-fiber IFU
    (MA.026 also shown in Fig.~\ref{fig:positional-errors}). The top
    number in each fiber maps to its position on the slit; the bottom
    number gives the lab-measured throughput (in percent). Sky fibers
    associated with the IFU are shown above and below the hexagonal
    array. \label{fig:ma026_tputhex}}
\end{figure}

To measure throughput, once the location of every fiber is known, the
2-axis stage at the input interface positions the input illumination
spot onto the center of every single fiber in a defined sequence. The
stages are very accurate, with uncertainties of $<1\mu$m for under 5
mm of travel, decreasing to $\sim10$nm on travel of order the fiber
diameter. As each fiber is illuminated, a second 2-axis stage
positions the output fiber blocks to peak up the emergent light signal
reimaged onto, and measured by, a photodiode in the heritage output
module. This signal is compared to the average signal from the direct
beam measured just before and just after the fiber measurements.
Sequences with direct-beam measurements that drift (before to after)
by more than 1\% are discarded, which rarely occurs when the lamp is
allowed to warm up for a period of one hour. A fiber harness is deemed
acceptable if all individual fiber throughput measurements are above
85\% before AR coating. In practice we measure average throughputs of
96\%$\pm$0.5\% after AR coating. The sequence also builds up a mapping
table between the IFU and slit. An example of the throughput output
plots can be seen in Fig.~\ref{fig:ma026_tputhex}.

To anticipate the comparison of laboratory to on-sky thoughput
measurements, several aspects of the SDSS test-stand are worth noting:

(1) The illumination spot is placed approximately at the center of
every fiber. Compared to on-telescope coupling, this near-field
illumination is more concentrated than the PSF in most seeing
conditions, and less diffuse than dome-flats or sky.  In principle
this might lead to differences in mode-coupling that would yield
throughput estimates in the lab unrepresentative of on-sky
performance.  However, we compared throughput measurments for
different spot locations on several fibers, and didn't not find
discernable throughput variations.

(2) The throughput is measured for a specific input spectrum defined
by the lamp temperature (3200~K at 100\% output) and a BG38 filter for
color-balance (located in the collimated beam of one-to-one reimaging
optics in the output module) convolved with the photo-diode response
function (Melles-Griot DSI 007). The lab measurements should be
representative of values measured on telescope in the $g$ and $r$
bands.

\begin{figure}[tbh]
  \centering
  \includegraphics[width=0.5\textwidth]{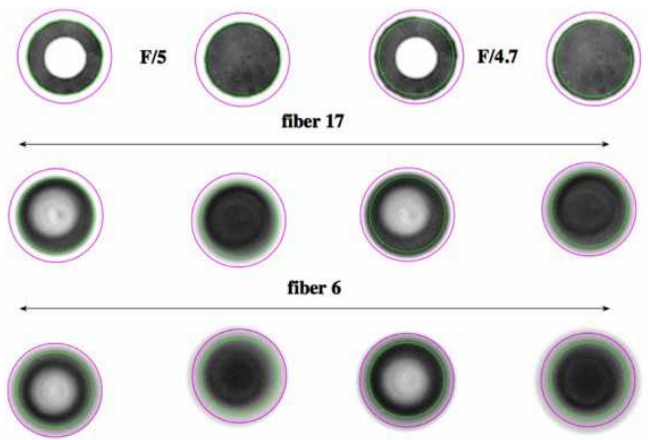}
  \caption{Far-field images of the input and output beams from TSW,
    designed to mimick differences between TS2 and the BOSS
    spectrograph. Far-field images images are for two fibers
    exhibiting the least and most FRD in harness MA023, as seen in
    Fig.~\ref{fig:tpc}. Green circles mark f/5 while magenta circles
    mark f/4.\label{fig:farfield}}
\end{figure}

\begin{figure}[tbh]
  \centering
  \includegraphics[width=8cm]{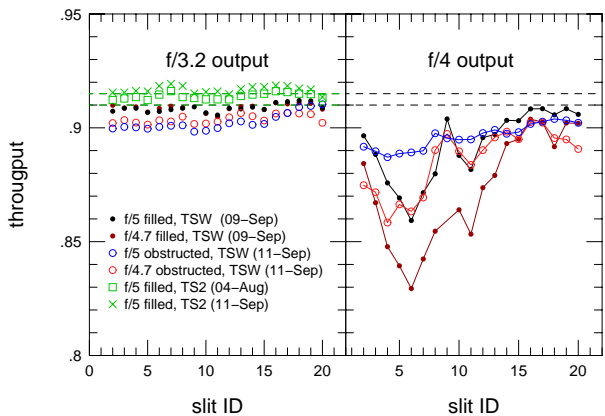}
  \caption{Throughput measured for the 19-fiber IFU harness MA053 (no
   AR coating) with TS2 and TSW.  For TSW, measurements are made for
   4 different input beams, illustrated in Fig.~\ref{fig:farfield}
   and encoded here in the key, and two output apertures
   corresponding to f/3.2 (left) and f/4 (right). The former mimicks
   the TS2 output module, while the latter mimicks the BOSS
   spectrograph optics. Throughput differences between TS2 and TSW
   for the same input beam are under 1\%.  Throughput differences
   between a filled, f/5 input beam and an f/3.2 output aperture with
   other input beams and output apertures are bewtween 1.3\% and
   3.5\% in the mean, with larger differences corresponding to faster
   and filled input beams.\label{fig:tpc}}
\end{figure}

(3) The SDSS input modules inject a highly uniform, f/5 input beam
into the fibers, matching the Sloan 2.5m telescope input f-ratio, but
missing the telescope's f/10 central obstruction.  Our lab
measurements also do not simulate the effects of non-telecentricity
introduced by drill errors, hole tolerances, and
plate-to-focal-surface mismatch (see
\S~\ref{sec:instrument-throughput} below). Non-telecentricity may
introduce an effective beam speed as fast as f/4.7. To investigate the
impact of differences between light injection in the lab and
on-telescope on system throughput, we re-measured the output beam
profiles for one 19-fiber harness (MA053, uncoated) on TSW with an
input pupil matching the Sloan 2.5m telescope (including the central
obstruction), and then again for an f/4.7 input beam (with the same
obstruction). Examples of the input and output far-field beams are
shown in Fig.~\ref{fig:farfield}.  The flux enclosed within f/4 output
beam is $89\pm2$\% for a filled f/5 input beam, rising only slightly
but with half the scatter to $90\pm1$\% when adding the central
obstruction, and returning to $89\pm2$\% for an f/4.7 input beam with
the same central obstruction.  In short, the detailed input beam
profile differences between the lab and on-telescope should have
marginal impact on performance estimates for light emerging from
fibers within f/4.

(4) We discovered that the output modules in TS1 and TS2 accept beam
speeds as fast as f/3.2 $\pm$ 0.03 (depending on up to $\pm$0.5~mm of
defocus). Faster light is rejected by use of a field stop, but this is
significantly faster than the f/4 acceptance of the BOSS spectrograph
optics. Based again on our measurements with TSW, (compared to TS2 in
Fig.~\ref{fig:tpc}), in the mean we find the percentage of light
coming out of the fiber between f/4 and f/3.2 is 2\% for a filled f/5
input beam; 1\% for an f/5 input beam including an f/10 central
obstruction; and 2\% for an f/4.7 input beam including an f/10 central
obstruction.  Some fibers with the most FRD exhibit two to three times
as much light at these fast output speeds, but in realistic conditions
on telescope, the difference between lab and telescope throughput
should be within 5\%.

\section{On-sky Performance}
\label{sec:on-sky-performance}

A primary constraint derived from the science requirements is that the
instrument perform as well or better than the single-fiber BOSS system
in both throughput and stability.  This is driven by simulations
conducted using BOSS data showing that BOSS-like performance would
meet MaNGA's objectives.

\subsection{AR Coating Gains}
\label{sec:ar-coating-gains}

The addition of an AR coating directly to the fiber faces on the IFU
and V-groove terminations has been shown to be effective in both
witness samples and laboratory testing of the fibers (see
\S~\ref{sec:ar-coating}).  A second useful check of the coatings
effectiveness was to test the transmission of the coated IFU fibers
next to the uncoated BOSS fibers to verify the improvement in
throughput.  A sequence of flat fields was taken with a BOSS
cartridge, and then while keeping the flat field lamps on, the same
sequence was repeated with a MaNGA cartridge.  To verify the stability
of the flat field lamps, another set of exposures was taken with the
BOSS cartridge after the MaNGA sequence.  Each flat was bias
subtracted and a pixel mask was applied.  A mean fiber throughput was
then generated for each flat field by summing up the total flux in the
image and dividing by the number of fibers on the CCD.  Over the
roughly 30 minutes of testing, the lamps themselves were remarkably
stable at less than 0.3\% variability.  The improvement seen from the
AR coating on the IFUs as compared to the BOSS fibers is $\sim$6\% in
the blue camera and $\sim$4\% in the red camera.  This is well in line
with what is expect from both reflectance curves of witness samples
and from results seen in the SDSS test stand which suggested a
$\sim$4\% gain in throughput.

\subsection{Detailed Throughput Measurements}
\label{sec:instrument-throughput}

The BOSS spectrographs accept light within an f/4 input cone from a
properly aligned fiber on the slit.  In an idealized instrument, the
only losses in total throughput are attenuation in the fiber,
110~dB/km at 300~nm (0.5\%; Fig.~\ref{fig:FBP-transmission}) and the
Fresnel losses at the ends of the fiber.  AR coating reduces these end
losses to $\sim$1.5\% (Fig.~\ref{fig:AR-reflectivity}), suggesting a
theoretical maximum throughput of 96.5\% through at an output of f/4.
Lab measurements (Section~\ref{sec:laboratory-measurements-of-ifus})
reveal a throughput value of 96$\pm$0.5\% into an f/3.2 aperture stop.
Lab testing accounts for FRD caused by surface finish at both input and
output, and stresses induced from epoxy at the ferrule and v-groove
ends of the fiber harness.

\begin{figure}[tb]
  \centering
  \includegraphics[width=8cm]{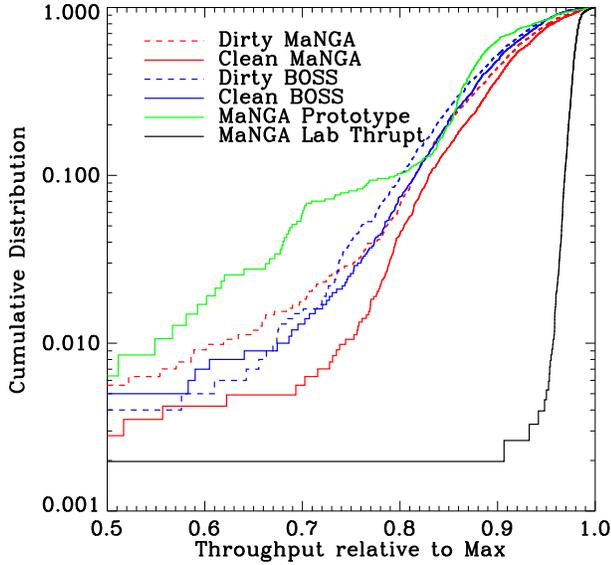}
  \caption{Histogram of relative throughput normalized to the highest
    performing fiber.  The dashed red line represents the throughput
    before cleaning, and the solid red line shows the performance
    improvements after cleaning.  The black line represents laboratory
    throughput values also normalized to the highs performing fiber.
    also included are the distribution of the BOSS single fiber system
    (Blue) and the MaNGA early prototype IFUs
    (Green).\label{fig:cleaning_hist}}
\end{figure}

The lab measurements do not account for stress induced FRD generated
from bending, twisting, or pulling the fibers within the cartridge,
input angular misalignment induced at the plug plate, misalignment at
the slit, or contamination-induced throughput losses from operating in
an imperfectly-controlled environment.  To fully understand the
throughput of the system in-use as compared to the laboratory
measurements (Fig.~\ref{fig:throughput_breakdown}), flat fields are
used to generate relative throughput plots of fibers with respect to
the best-performing fiber within a given cartridge.
Figure~\ref{fig:cleaning_hist} shows the fiber throughput distribution
of fibers within a single cartridge.  All of the single traces within
this plot are normalized to the highest performing fiber within that
set of flats, so this plot is only useful in understanding the
throughput distribution within a cartridge and can not be used as a
absolute comparison between lab measurements and on telescope
measurements.

It is immediately obvious from Fig.~\ref{fig:cleaning_hist} that there
is a difference between the distribution of throughput values measured
in the lab (solid black line), where 90\% of the fibers achieve
$\sim$96\% of the throughput of the highest performing fiber, and the
on-telescope measurement (solid red line) which suggests that 90\% of
the fibers perform within 83\% of the best case. This significantly
higher level of throughput variation is attributed to slightly higher
FRD on the telescope (see Section~\ref{sec:frd}), compounding beam
pointing errors, vignetting in the spectrographs, residue
contamination of the IFUs, and other unknown sources of losses. The
breakdown of these additional losses is discussed in detaol below. It
is worth pointing out that when the IFUs are clean, the level of
throughtput variation between fibers is significantly lower than that
found in clean BOSS single fiber ferrules
(Fig.~\ref{fig:cleaning_hist}). When allowed to become contaminated
the throughput variation between fibers becomes comparable to that
found in the BOSS single fiber system. Cleaning methods discussed in
section~\ref{sec:ifu-contamination} reduce the level of contamination
on the IFUs.  The on-telescope performance of both the BOSS and MaNGA
fibers has far less variability than the early prototype
IFUs. Analysis of the prototype data clearly led to significant
improvements in virtually all components of the MaNGA system.

\subsubsection{Stress-Induced Focal Radio Degradation}
\label{sec:frd}

\begin{figure}[tb]
  \centering
  \includegraphics[width=0.4\textwidth]{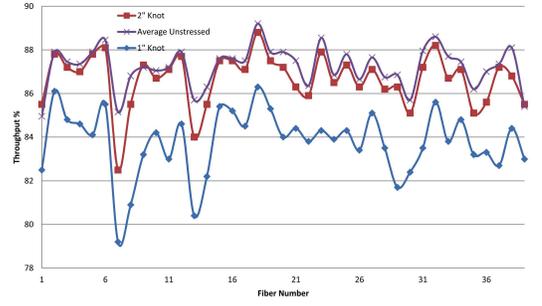}
  \caption{Effect of stress on a 37 fiber IFU tested on the SDSS test
    stand.  A smaller aperture stop was used to help capture
    variations from stress.  The purple line marked with X's indicates
    the unstresses state of the harness.  Red squares shows the
    stress induced under a 2 inch diameter bend, and the blue
    represents throughput seen under a 1 inch diameter bent.  A 2 inch
    bend diameter is possible but difficult to achieve on the
    telescope while a 1 inch bend radius is the physical limit of the
    fiber jacketing and would not be seen in operations.  In general
    no more than $\sim$4\% throughput effect is observed from FRD
    induced under bending stress in the
    fibers.\label{fig:frd_labplot_V01}}
\end{figure}

\begin{figure}[tb]
  \centering
  \includegraphics[width=0.4\textwidth]{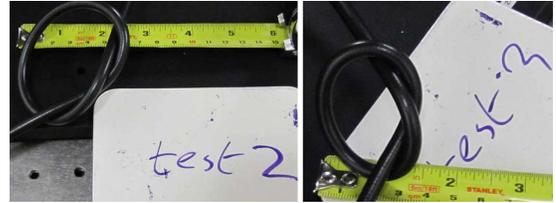}
  \caption{Right: Image of the 37 fiber IFU bent to a 2 inch diameter.
    Right: Image of the 37 fiber IFU bent to a 1 inch diameter.  Lab
    throughput measurements were taken under both stress states to
    determine how bending within the cartridge no the telescope would
    affect throughput.\label{fig:frd_image_V01}}
\end{figure}

In an attempt to quantify the stress-induced FRD that could be
contributing to the higher variability in fiber throughput in the
cartridge, an IFU harness in the lab was subjected to bending, and
torsional loading in excess of what would be experienced in operation
at the telescope. Typical bend radii within a cartridge are seen to be
no smaller $\sim$2~inches in radius.  A single bend within this stress
state will induce an average FRD loss of $\sim$0.6\% with an RMS at
about the same level, the maximum loss seen was 2.65\%.  A more
extreme bending case was evaluated to understand the upper limit in
FRD a bend in the IFU cable could produce.  This was done by tying a
knot in the plugging section of the IFU cable
(Fig.~\ref{fig:frd_image_V01}) with as small as a 0.5~inch bend
radius.  This exceedingly tight bend produced significantly higher
losses of $\sim$3.5\% on average with a maximum observed loss of
$\sim$6\%. Fig.~\ref{fig:frd_labplot_V01} showes the throughput for
all 39 fibers in the harness used for this test under 0.5 and 1 inch
bend radius.  Torsional loading had less of an effect on the
throughput, a 360 degree rotation produced on average a $\sim$1\% loss
in throughput with a maximum loss of 1.6\%.  Somewhat surprisingly,
torsional bends of 180 and 720 degrees produced similar averages and
maximum throughput losses as the 360 degree test.

Given the layout of the cartridge and the positions the fibers are put
into when plugged, it is conceivable to produce a few bends and two
twists in the harnesses. Using these tests as a guide we estimate a
contribution of typically $\sim3$\% and a maximum of $\sim$6\% to the
total difference between lab and on telescope throughput.

\begin{figure}[tb]
  \centering
  \includegraphics[width=7cm]{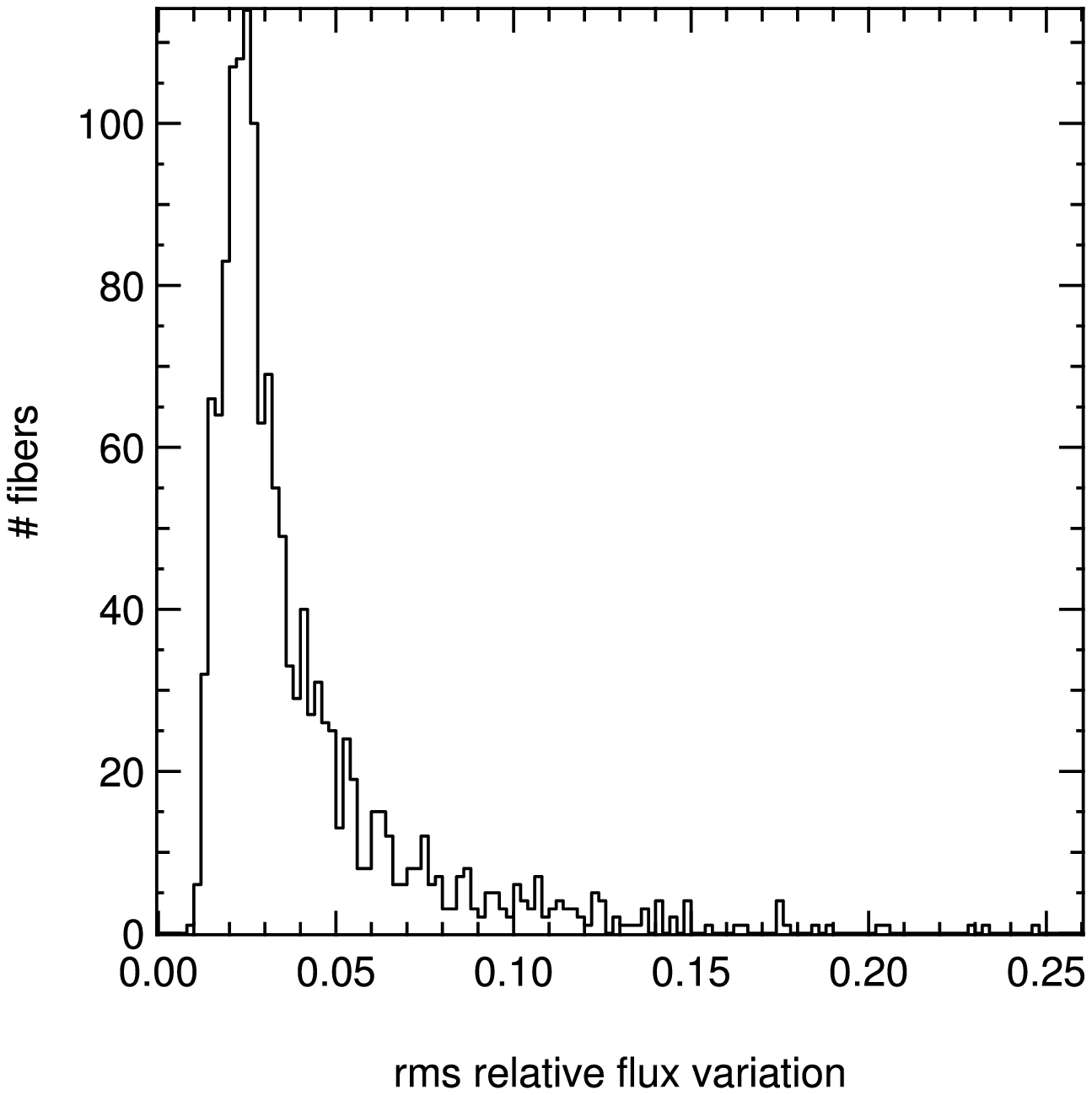}
  \includegraphics[width=7cm]{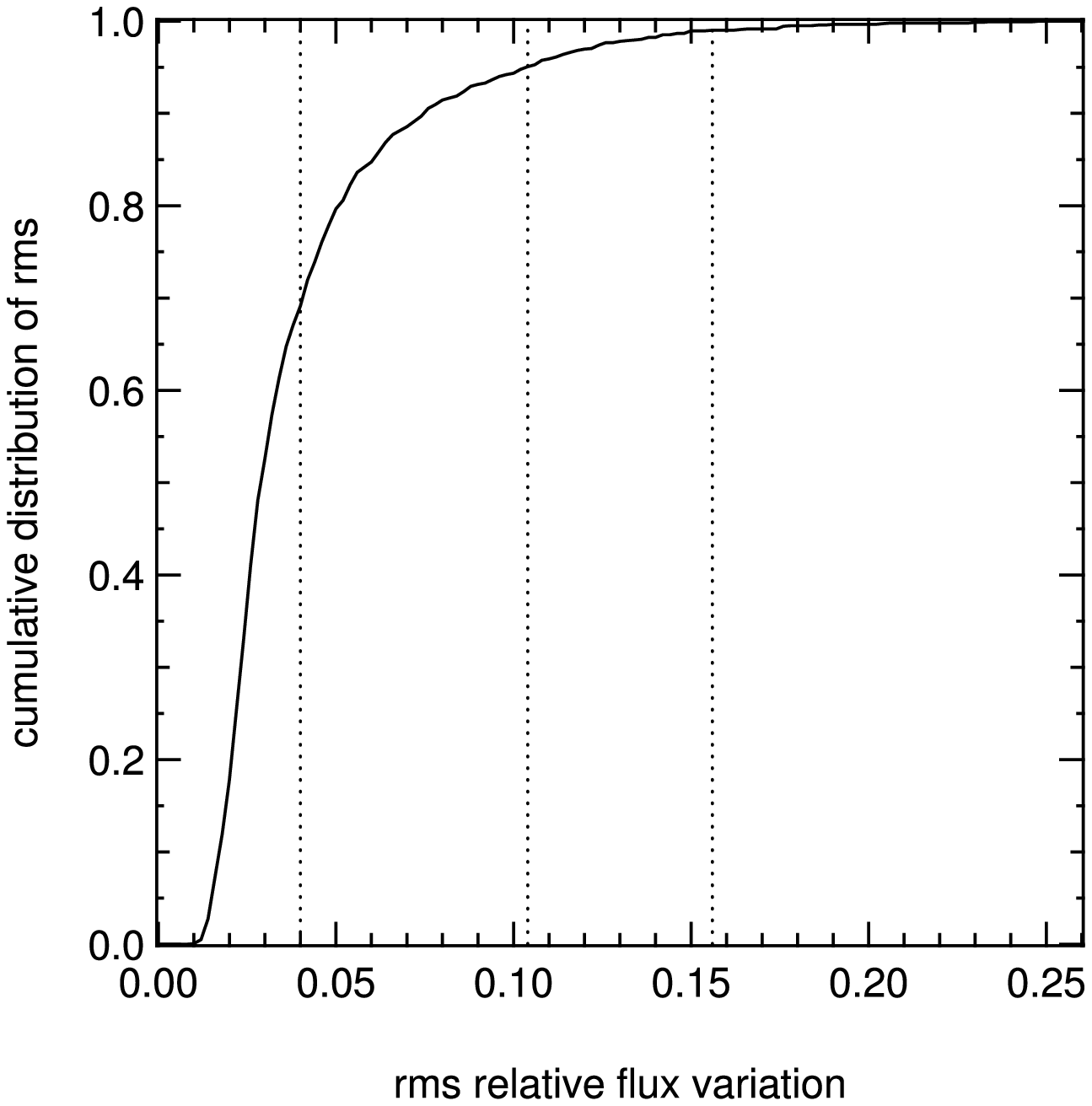}
  \caption{Variation of fiber throughput from plugging to plugging
    (differential, top panel; cumulative, bottom panel). The
    variablity (in RMS sense) of each fiber is measured in a sequnce
    of 13 flat fields with a change of plate and replugging in between
    each.\label{fig:fiber_variation_plugging}}
\end{figure}

An independent test of the conjecture that stress-induced FRD in
plugging contributes to the throughput losses can be obtained by
realizing that these losses would be variable from plugging to
plugging, while other sources of losses are static (at least on short
time scales). Hence we look at the {\em relative variablity} of each
fiber in sequences of flat fields taken with the same plate, but with
the fibers removed and replugged in between each flat exposure (13
exposures total). The results are shown in
Fig.~\ref{fig:fiber_variation_plugging}. In quite good agreement to
the lab analysis, we find that the mode of the distribuiton of
variability is at 2.5\% with the 68th percentile at 4\%. The effect of
this stress-induced component is shown as the magenta line in
Fig.~\ref{fig:throughput_breakdown}.

\subsubsection{Slit Misalignment}

While stress-induced FRD and vignetting does explain most of the
difference in throughput between the lab measurements and the
on-telescope measurements, it dose not explain all of the difference
(see Fig.~\ref{fig:throughput_breakdown}).  Originally, beam angle
both at the telescope focal plane and the spectrograph input was
considered a prime suspect for the remaining variability.  After
further investigation this proved to not be the case.  Variability on
the slit plate angle seen from dummy builds as well as measured
variations of production components show a maximum angular
misalignment of 0.003 radians or 0.17 degrees (see
\ref{sec:v-groove-block-alignment}). This alignment precision is
confirmed by comparing the flux in arc line images taken with one of
the two Hartman doors installed in the spectrographs closed. The
difference in flux between the left Hartmann and right Hartmann image
of the same line corresponds to first order to a pupil illumination
difference due to misalignmet. These data suggest that the fibers are
aimed correctly at the 0.14 degree level, in line with the mechanical
measurements. If we assume a uniform beam profile coming out of the
fibers at f/4 this will translate into a less than 0.1\% throughput
loss.

\subsubsection{IFU Ferrule/Plate Misalignment}

The IFUs can show up to 0.35 degrees of angular misalignment with
respect to the chief ray at the telescope focal plane due to
mechanical tolerances of the plate holes and ferrules (see
\S~\ref{sec:plug-plage-interface}). When plugged, forces from the bent
sheathing very likely push the ferrules from the neutral position to
the extreme of motion allowed by the slightly oversized hole.  When
the ferrule and hole are at the extreme of the fabrication tolerance
range, this will cause degradation of the output beam from f/5 to
f/4.7 in the absence of any other sources of fiber-induced FRD simply
because the fiber is a perfect azimuthal scrambler of the input beam.
This geometrically induced faster output beam is often referred to as
geometric FRD, but is exactly what is described by \citet{Wynne1993}
as non-telecentric FRD.  At the fiber output, in our case at most 6\%
of the light is pushed beyond the f/4 acceptance cone of the
collimator (\S~\ref{sec:laboratory-measurements-of-ifus}; see also
\citet{Murphy+2008}). We cannot measure this effect directly in our
hardware or data. We therefore simulate plug plates by randomly
choosing ferrule-hole clearances using the measured ferrule and hole
tolerances (Tables~\ref{instrument:tab:ferrule_tolerance} and
\ref{instrument:tab:plugging_error}), calculating the resulting tilt
of the ferrule, and using the measurements from
\S~\ref{sec:laboratory-measurements-of-ifus} and Fig.~\ref{fig:tpc} to
estimate the amount of light lost for each IFU. We generate 25 such
realizations and add them fiber by fiber to the cumulative throughput
distributions shown in Fig.~\ref{fig:throughput_breakdown}. The range
of results (peak-to-peak) is shown as the green shaded region which
seems to lead to throughput values consistent with what we measure on
the telescope (red solid line).  Hence this effect seems to account
for all of the remaining throughput losses.

\subsubsection{Stability with Flexing and Gravity}
\label{sec:stability}

MaNGA observations are divided into sets of 3 dithered 15-minute
exposures.  Under optimal observing conditions a single plate can
reach the required S/N in two such dither sets, however, needing three
exposure sets will be the common case.  Because of constraints imposed
by differential atmospheric refraction (DAR) on the hour angle range a
plate can be observed at \citep{Law+2014a}, the maximum time between
calibration exposures is $\sim$90 minutes (a full description of MaNGA
observing is given in \citet{Yan+2014} and \citet{Law+2014a}.)

\begin{figure}[tb]
  \centering
  \includegraphics[width=8cm]{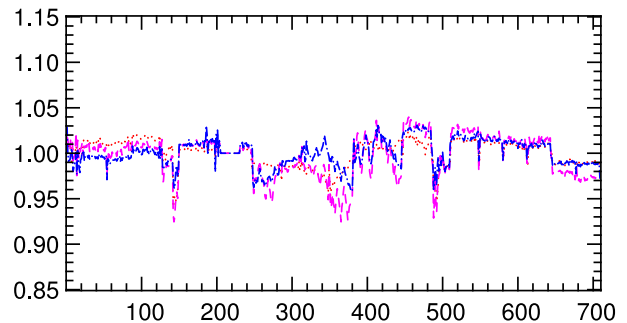}
  \caption{Fiber throughput normalized to the highest performing fiber
    vs the fiber ID.  Each trace illustrates a different position of
    the field rotator of the 2.5m telescope which corresponds directly
    to a change in gravity vector.  Maximum variability seen over 360
    degrees of rotation $\sim$3\%, in practice over a given set of
    exposures the gravity vector will only lead to 0.25\% change in
    instrument throughput (measured from before to after exposure
    calibrations).\label{fig:throughput_stability}}
\end{figure}

Because the IFU cables are significantly longer and heavier than the
BOSS single fiber ferrules, there is concern over the potential for
the changing gravity load during an exposure set due to rotator motion
to cause throughput variability by inducing FRD in the fibers,
instrument and telescope flexure, or possibly movements of the fibers
themselves within their plugging hole.  None of these causes could be
easily simulated in the lab so tests were conducted on the telescope
to characterize variability induced by changes in the gravity vector.
With the telescope at the stow altitude of $\sim$30 degrees, we rotate
the instrument through 270 degrees and take flat fields at four
positions. Figure~\ref{fig:throughput_stability} shows the variability
in throughput seen in this test.  Not surprisingly, the largest effect
on throughput can be seen at a 180 degree rotation.  The variability
over this large of a change in rotator angle can be as high as
$\sim$3\%.  In practise, however, over a 90 minute observing window
the maximum change in the gravity vector on the instrument is no
larger than $\sim$19 degrees, which leads to less than 0.25\%
variability in throughput.

\subsubsection{IFU Contamination}
\label{sec:ifu-contamination}

\begin{figure}[tb]
  \centering
  \includegraphics[width=0.4\textwidth]{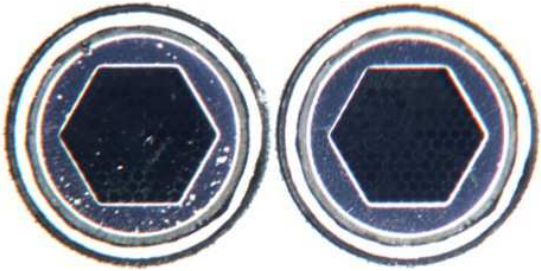}
  \caption{Left: Image of 127-fiber IFU Ma.023 with contamination from
    plugging operations.  Aluminum particles can be seen in the center
    and lower left of the IFU as well as a haze across the lower right
    caused by finger oil contamination from the plugging operation.
    Right: The same 127 fiber IFU cleaned with pure
    methanol. \label{fig:ma23_cleaning}}
\end{figure}

Contamination of the IFU fiber faces is also seen to be a significant
effect on overall performance of the fiber system.  In particular,
extremely low performing fibers ($<$80\% throughput compared to the
best fiber) are almost always attributable to contamination on the IFU
face.  While the site itself can be extremely dusty, particularly in
the spring, the majority of the contamination of the MaNGA IFUs come
from handling during plugging operations.  Particulate contamination
in the form of aluminum dust generated from the plug plate and
environmental contamination such as pollen and gypsum have been
observed on the faces of the IFUs (Fig.~\ref{fig:ma23_cleaning}).
Chemical contamination from finger oils and machine grease found on
the cartridge handling system have also been found to be present and
affect the amount and rate at which particles adhere to the IFU faces.

Contamination has been observed to build up after only a few pluggings
and can cause a significant loss in throughput both in localized areas
of the IFU leading to higher variability, and across the entire bundle
leading to degraded overall throughput at about the $\sim$10\% level
(see before/after cleaning curves in Fig.~\ref{fig:cleaning_hist}).

In order to reduce throughput losses due to contamination a number of
mitigation strategies aiming at better environmental control have been
put in place.  Pluggers are now required to wear cleanroom gloves when
handling the fibers.  Grease and dust contamination from equipment has
been reduced through regular wipedowns of the equipment and the
plugging lab.  CO2 snow cleaning of the each plate after plugging is
complete has been introduced to remove any dust prior to stowing the
(covered) cartridge in the bay for later observations.  Lastly,
ultrasonic cleaning for one minute with immersion in pure methanol has
been shown to remove all contamination from the surface and returns
the IFUs to pristine condition. This procedure is now performed
monthly on all fibers.

\subsection{Throughput Summary}

\begin{figure}[tb]
  \centering
  \includegraphics[width=0.4\textwidth]{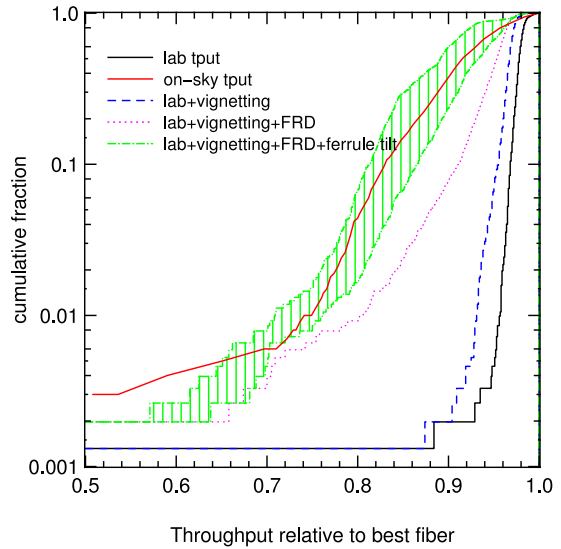}
  \caption{Breakdown of throughput losses due to various
    causes discussed in the text. The solid black line shows the
    cumulative throughput distribution measured in the lab
    (Fig.~\ref{fig:cleaning_hist}) The solid red line is the
    distribution measured at the telescope taken from the same figure.
    The dashed blue line adds the vignetting by the
    spectrograph+camera optics as a function of slit position. The
    dotted magenta line represents the lab+vignetting distribution
    convolved with the plugging-stess FRD losses according to
    Fig.~\ref{fig:fiber_variation_plugging}. The dash-dotted green
    line adds the geometric FRD effect caused by the tilt of the
    ferrule within the plug hole, using the measured hole/ferrule
    tolerances.\label{fig:throughput_breakdown}}
\end{figure}

We show the breakdown of throughput losses due to various causes
discussed in the sections above in
Fig.~\ref{fig:throughput_breakdown}. We take the throughput
distribution of our fibers as measured in the lab and on-telescope
(see Fig.~\ref{fig:cleaning_hist}) and compare it to the sum of FRD
losses (Fig.~\ref{fig:fiber_variation_plugging}), geometric FRD
effects, and losses due to vignetting along the slit. We note that
these effects explain the bulk of the losses seen between the lab
measurements and the on-telescope data. Additional effects, such as
fiber surface contamination (after cleaning), fiber/slit alignment
issues, flexing in the spectrograph, must contribute only of the order
of a few percent at most.  All effects discussed in the above sections
are equally affecting the MaNGA IFUs and the single-fiber BOSS system.

\section{Summary and Outlook}
\label{sec:conclusions}

We describe the design, manufacture, and performance of bare-fiber
integral field units for the MaNGA survey (Mapping Nearby Galaxies at
APO) within SDSS-IV at the Sloan 2.5~m telescope at Apache Point
Observatory (APO). Our IFUs have hexagonal dense packing of fibers
with packing regularity of 3~$\mu$m (RMS), and throughput of
96$\pm$0.5\% from 350~nm to 1~$\mu$m in the lab. The IFUs for MaNGA
range from 19 to 127 fibers (3-7 hexagonal layers). High throughput
(and low focal-ratio degradation) is achieved via intact cladding and
buffer, excellent surface polish, and direct application of
multi-layer AR coating to the input (IFU) and output (slit) fiber
surfaces. On-sky data illustrate the performance of this IFU
design. In operations at Apache Point Observatory, the IFUs show only
an additional 2.5\% FRD-related variability in throughput despite
repeated mechanical stressing during plugging and unplugging of
plates, and achieve on-sky throughput 5\% above the single-fiber feeds
used in SDSS-III/BOSS, attributable to no worse FRD and the addition
of the AR coating.  The manufacturing process is geared toward
mass-production of high-multiplex systems. The low-stress process
involves a precision ferrule with hexagonal inner surface shape
designed to lead inserted fibers to settle in a near-ideal dense
hexagonal pattern. The ferrule inner diameter is tapered at
progressively shallower angles toward its tip and the final 2~mm are
straight and only a few micron larger than necessary to hold the
desired number of fibers. This ferrule is made possible by the
application of EDM fabrication processes. To assure quality, automated
testing in a simple and inexpensive system enables complete
characterization of throughput and fiber metrology. The IFU
fabrication process described here uses 120:132:150~$\mu$m
core:clad:buffer fibers to achieve filling factors of 56\%, but can be
applied to other fiber sizes and larger numbers of fibers.  Future
applications of this IFU manufacturing process include much larger
IFUs, and possibly higher fill-factors with stripped buffer,
de-cladding, or lenslet coupling.

It is of broader interest to reflect here on several salient features
of our design.  The first concerns the trade-off between
filling-factor and field coverage in the context of a specific survey
sample and science goals.  To cover a certain number of galaxies over
a range of angular scales depends primarily on the total number of
spatial apertures (e.g., fibers) of a given size, not how tightly they
are configured. What drives our filling factor is, on one hand, the
desire to maintain fidelity in the spatial sampling and reconstruction
of the spectral image. On the other hand, we must match the angular
scale of our IFUs with the surface-density of suitable targets. It is
simply fortuitous that the fiber spacing required for
image-reconstruction fidelity yields a good match to the telescope and
spectrograph fields of view and the available targets; we have simply
capitalized on this happenstance.

Directly related to fiber spacing scale is the consideration of
spacing regularity. As we have shown, high-precision placement of
fibers in a regular grid enables accurate image reconstruction without
100\% fill-factor. In particular, this is the case even in the
presence of differential atmospheric refraction, where for for
wide-field survey applications atmospheric dispersion correctors are
impractical or cost-ineffective. Our technical achievement in
high-precision fiber IFUs will help future instruments interface with
lenslet arrays and minimize A$\Omega$\ losses (entropy
increase). These losses occur due to the necessity of over-sizing
fibers to accommodate for alignment errors between the fiber input face
and the pupil image formed by the lenslet, and the ensuant radial
scrambling over the fiber run which fills the output fiber face.

\acknowledgements

ND acknowledges support through CONACyT grants 167332 and 180125, and
the hospitality of the Max-Planck Institute for Extraterrestrial
Physics during the writing of this manuscript. KB was supported by
World Premier International Research Center Initiative (WPI
Initiative), MEXT, Japan. MAB acknowledges the hospitality of the
Institute for Cosmology and Gravitation (Portsmouth University), and
support from the Leverhulme Foundation.

We are grateful to J.~Gonz\'alez, J.~Bland Hawthorne, G.~J.\ Hill,
S.~S\'anchez, F.~Hearty, R.~Sharples, M.~Blanton, C.~Rockosi, M.~Roth,
and A.~Pecontal for advice, reviews, and helpdul discussions of all
topics surrounding fibers, IFUs, and SDSS. We are grateful to
J.~Etherington, K.~Masters, R.~McDermid, O.~Steele, K.~Thanjavur,
D.~Thomas, D.~Wilkinson, T.~Xiao, and K.~Zhang for analysis of and
insights from data collected with prototype IFUs.

Funding for the Sloan Digital Sky Survey IV has been provided by the
Alfred P. Sloan Foundation and the Participating Institutions. SDSS-IV
acknowledges support and resources from the Center for
High-Performance Computing at the University of Utah. The SDSS web
site is www.sdss.org.

SDSS-IV is managed by the Astrophysical Research Consortium for the
Participating Institutions of the SDSS Collaboration including the
Carnegie Institution for Science, Carnegie Mellon University, the
Chilean Participation Group, Harvard-Smithsonian Center for
Astrophysics, Instituto de Astrof\'isica de Canarias, The Johns Hopkins
University, Kavli Institute for the Physics and Mathematics of the
Universe (IPMU) / University of Tokyo, Lawrence Berkeley National
Laboratory, Leibniz Institut f\"ur Astrophysik Potsdam (AIP),
Max-Planck-Institut f\"ur Astrophysik (MPA Garching),
Max-Planck-Institut f\"ur Extraterrestrische Physik (MPE),
Max-Planck-Institut f\"ur Astronomie (MPIA Heidelberg), National
Astronomical Observatory of China, New Mexico State University, New
York University, The Ohio State University, Pennsylvania State
University, Shanghai Astronomical Observatory, United Kingdom
Participation Group, Universidad Nacional Aut\'onoma de M\'exico,
University of Arizona, University of Colorado Boulder, University of
Portsmouth, University of Utah, University of Washington, University
of Wisconsin, Vanderbilt University, and Yale University.


\end{document}